\documentclass[a4paper,11pt]{scrartcl}

\usepackage{afterpage}

\usepackage{setspace}

\usepackage{fix-cm}
\usepackage{graphicx}
\usepackage{geometry}
\usepackage[authoryear]{natbib}
\usepackage[thinspace,thinqspace,squaren,textstyle]{SIunits}		
\addunit{\yr}{a}
\addunit{\sv}{Sv}
\addunit{\psu}{psu}
\addunit{\ppm}{ppm}

\usepackage{subfig}		

\usepackage{booktabs}		

\usepackage{longtable}

\usepackage{color}		
\usepackage{tikz}		
\usepackage{pgfplots}
\usetikzlibrary{decorations}
\usetikzlibrary{mindmap}
\usetikzlibrary{trees}
\usetikzlibrary{calc}
\usetikzlibrary{arrows}
\usetikzlibrary{plotmarks}
\usetikzlibrary{positioning}
\usetikzlibrary{fadings}
\usetikzlibrary{decorations.pathmorphing}
\usetikzlibrary{decorations.markings}
\usetikzlibrary{decorations.pathreplacing}
\usetikzlibrary{shapes,shadows}
\usepgflibrary{arrows}

\definecolor{mygray}{HTML}{969696}
\definecolor{aqua}{HTML}{00BFFF} 

\newcommand{\CO}{CO$_2$}
\newcommand{\DP}{\textit{Drake Passage}} 
\newcommand{\contr}{\textit{control}}
\newcommand{\final}{\textit{final}}
\newcommand{\COtest}{\textit{CO$_\mathit{2}$-test}}
\newcommand{\Stestelf}{\textit{S-test-11.5}}
\newcommand{\Stestzz}{\textit{S-test-22.5}}
\newcommand{\Stest}{\textit{S-test}}
\newcommand{\DPtest}{\textit{DP-test}}

\usepackage[affil-it]{authblk}

\title{The role of atmospheric greenhouse gases, orbital parameters, and southern ocean gateways: an idealized model study}

\author[1]{Eileen Hertwig
  \thanks{Max Planck Institute for Meteorology, Bundesstra\ss e 53, 20146 Hamburg, Germany, eileen.hertwig@mpimet.mpg.de}}        
  \affil[1]{Max Planck Institute for Meteorology, Hamburg, Germany}

\author[2]{Frank Lunkeit}        
  \affil[2]{Meteorological Institute, University of Hamburg, Germany}

\author[1,2]{Klaus Fraedrich}

\date{}

\begin{document}

\begin{titlepage}

 \maketitle

\begin{abstract}

A set of idealized experiments are performed to analyze the competing effects of declining atmospheric \CO{} concentrations, the opening of an ocean gateway, and varying orbital parameters. 
These forcing mechanisms, which influence the global mean climate state, may have played a role for triggering climate transitions of the past (for example during the Eocene--Oligocene climate transition and the build-up of the Antarctic Ice Sheet). 
Sensitivity simulations with a coupled atmosphere--ocean general circulation model are conducted to test these three forcings and their roles for the global climate. 
The simulations are carried out under idealized conditions to focus on the essentials. 
The combination of all three forcings triggers a climate transition which resembles the onset of the Antarctic glaciation. 
In particular, the temperatures in the southern high latitudes decrease and snow accumulates constantly. 
Moreover, the relative importance of each possible forcing is explored. 
All three of the mechanisms (atmospheric \CO{} decrease, opening of the ocean gateways, and changing orbital parameters) cool the climate of the southern polar continent. 
The change induced by the orbital parameters depends strongly on the magnitude of the change in obliquity.  

\end{abstract}

\end{titlepage}

\section{Introduction}
\label{sec:intro}

In climate science the methods that are available to observe or to model the atmosphere and the ocean are continuously improving. 
Since the introduction of satellite remote sensing into climate research, an entirely new spectrum of possibilites has opened up and many phenomena have been discovered or documented since. 
Furthermore, over the past decades there has been huge progress in climate modeling and general circulation models (GCMs) have benefited greatly from enhanced computational power. 
While model formulations and parameterizations are becoming more sophisticated, the increasing computational resources allow a finer resolution, faster (and therefore longer) simulations, and the explicit inclusion of more differentiated processes. 
Hence, climate simulations increasingly improve. 

In spite of the considerable progress already achieved in the research of climate dynamics, fundamental questions about our climate system remain that have not been answered up to this day. 
Climate changes of the past, for example, are not well understood. 
There is ample evidence from paleorecords that the climate system is not only slowly varying but may also undergo sudden shifts from one regime to another \citep{Broecker1985,Dansgaard1993,Jouzel1995}. 
However, what sets the time scales of these abrupt climate changes and what causes them is still under discussion. 
The role of the ocean, more specificly the thermohaline circulation (THC), has often been studied in this context \citep[e.g.][]{Winton1993,Rahmstorf1995}, but a definite answer has not been given. 

A prominent example of past climate changes is the glaciation of Antarctica, which occurred about 34 million years ago \citep[near the Eocene--Oligocene transition, see][]{Zachos2008} and for which the main forcing mechanism remains to be determined. 
The opening of the Drake Passage has often been in the center of discussion \citep[e.g.][]{Mikolajewicz1993,Toggweiler2000,Sijp2004,Cristini2012}. 
The opening of seaways permitted unrestricted circumpolar flow: the Antarctic Circumpolar Current (ACC) developed, which progressively thermally isolated Antarctica \citep[e.g.][]{Kennett1977,Kennett1978}. 
The ACC is a powerful current that stretches over a length of approximately \unit{20,000}\kilo\meter{} and extends down to a depth of \unit{2,000}\meter{}. 
It is the only current to completely encircle the globe linking the three main ocean basins (Atlantic, Pacific, and Indic) into one global system by transporting heat and salt \citep{Turner2009}. 
If and how the opening of an ocean gateway has influenced the onset of the Antarctic glaciation has not been conclusively answered. 
More generally, how topographic contraints on the ocean circulation impact the global climate or have influenced the climate during past eras is still an open question. 

The role of changing atmospheric greenhouse gas concentrations is often discussed in the context of past climate changes \citep[e.g.][]{Beerling2002,Hegerl2003,Winguth2010}. 
Furthermore, a change in atmospheric \CO{} might explain the onset of the Antarctic glaciation \citep{DeConto2003a,DeConto2003}. 
Atmospheric \CO{} concentrations were much higher before the Eocene--Oligocene transition than today. 
Declining greenhouse gases could have caused the atmospheric temperatures to decrease, which would have permitted substantial ice formation on Antarctica. 
However, there is still no consensus on the degree that changing atmospheric \CO{} has influenced the climate during past eras. 

Periodic changes in the parameters of the Earth's orbit around the sun affect the amount and distribution of incoming solar radiation and explain the periodic appearance of ice ages \citep{Milankovitch1941}. 
Milankovitch identifies three types of orbital variations, which could act as climate forcing mechanisms: obliquity of the Earth's axis, eccentricity of the Earth's orbit around the Sun, and precession of the equinoxes. 
The Eocene--Oligocene transition coincides with an orbital configuration comprising a phase of low eccentricity and low-amplitude change in obliquity, which dampens the seasonal cycle \citep{Coxall2005}. 
Thus, the glaciation of the Antarctic continent could also have been triggered by the configuration of Earth's orbital parameters, which favored ice sheet growth. 
The role of the orbital parameters during the Eocene--Oligocene climate transition and, more generally, how varying orbital parameters influence the global climate state is still a topic of research. 

There are different approaches to address these (and many more) open questions in climate research. 
One possibility is to achieve higher levels of complexity and resolution in climate models or to further improve observational techniques. 
A more detailed description of the climate state can indeed help to better explain climate phenomena and to make better predictions. 
However, another approach refrains from applying increasingly complex techniques and does not aim at describing the climate state as realistically as possible, but is rather directed at understanding the underlying mechanisms. 
If the model is confined to the most elemental processes, fundamental characteristics of the coupled climate system become easier to identify. 
The philosophy behind this method is to first understand the basic and most fundamental processes, before including more complex phenomena into the model of the climate system. 
This is the approach chosen in this study by means of applying an aquaplanet concept. 
An aquaplanet is a planet, where the entire surface of the Earth is covered with an ocean with a flat bottom and without any geometrical constraints. 
Aquaplanets can be a very useful tool for analyzing the climate state in its most elemental form, representing only a crude approximation of the Earth but with the same governing processes. 

This study aims at analyzing the roles of topographic constraints on the ocean circulation, of declining atmospheric \CO{}, and of changing orbital parameters for the global climate. 
In past studies, these three mechanisms have been identified as the main forcings for the onset of the Antarctic glaciation. 
For example, \citet{Mikolajewicz1993} study the effect of Drake and Panama gateways on the ocean circulation and conclude that the temperature changes due to an opening of the Drake Passage are not large enough to have triggered the Antarctic glaciation. 
They suggest that some other mechanism (possibly \CO{}) may be responsible for the expansion of the Antarctic Ice Sheet (AIS). 
\citet{Toggweiler2000} use an idealized water planet model to examine the effect of the opening of the Drake Passage on Earth's climate and find that it cools the high latitudes of the southern hemisphere. 
They conclude that the effect of the Drake Passage opening is greater than previously supposed. 
\citet{DeConto2003a,DeConto2003} find that the opening of the Drake Passage and the changing orbital parameters have a smaller effect on the southern hemisphere climate, relative to changes in atmospheric \CO{}. 
\citet{Cristini2012} conducted sensitivity experiments (opened and closed Drake Passage) and find that the AIS is larger with a fully developed ACC. 
They conclude that the opening of the Drake Passage contributed substantially to the Antarctic glaciation. 

The onset of the Antarctic glaciation is the motivation for choosing the three specific forcings in this study. 
However, an idealized model is applied and it is emphasized that the purpose of this work is not to reconstruct the Eocene--Oligocene climate transition, but to analyze the opening of ocean gateways, decreasing \CO{} concentrations, and changing orbital parameters in a more general manner to study their roles for the global climate system. 
Studying the basic processes that are related to these forcing mechanisms may contribute to a better understanding of past climate transitions. 
In contrast to \citet{DeConto2003a,DeConto2003}, who also study these three forcings and their roles for the onset of the Antarctic glaciation with a complex GCM, we apply a simpler model to limit the number of feedbacks and systematically analyze each of the three forcing mechanisms. 

A coupled atmosphere--ocean--sea ice model is applied for idealized numerical sensitivity simulations. 
To focus on the essentials an expanded aquaplanet set-up is applied by including two topographical features in a planet covered by a single ocean: a circular continent at the south pole and a meridional barrier in the ocean. 
The latter allows to mimic a closed or an open ocean gateway. 
Simulations are carried out with different atmospheric \CO{} levels and orbital parameters. 

There are previous studies which apply the general concept of an aquaplanet with some topography included to analyze the ocean circulation in relevance to orographic barriers. 
A general circulation model, which comprises two identical idealized ocean basins that are connected by a circumpolar channel at the south pole (representing the Atlantic and Pacific oceans), is used by \citet{Marotzke1991} who find multiple steady states of the ocean, which are characterized by deep-water formation in different basins. 
\citet{Toggweiler2000} applied a water planet model with an opened and a closed meridional barrier in the southern ocean (similar to the one applied in this study) and observe a strong cooling effect in the southern hemisphere when the passage is opened, which is caused by a cross-equatorial overturning circulation and westerly winds over the open channel.
However, the simulations of \citet{Toggweiler2000} do not include a continent at the south pole but two small polar islands, which are not treated as land-surface (i.e.\ which do not include snow cover or ice--albedo feedback). 
\citet{Smith2006} have been the first to apply a fully coupled atmosphere--ocean general circulation model to an aquaplanet.  
They analyze the large-scale ocean circulation in aquaplanets with meridional ocean barriers and find that even though changes in the land--sea configuration cause regional climate deviations, there is hardly an impact on heat transports within the system.
\citet{Enderton2009} use similar aquaplanet set-ups to explore to which degree the total meridional heat transport is sensitive to the details of its atmospheric and oceanic components. 
Completely different climate states are simulated, of which the solution with an idealized Drake Passage shows the closest resemblance to our present-day climate. 

The model used in this study and the design of the sensitivity experiments are introduced in section~\ref{sec:model}. 
The results are discussed in section~\ref{sec:results}. 
A summary and conclusion is given in section~\ref{sec:conclusion} together with a short outlook. 

\section{The coupled model}
\label{sec:model}

The numerical model applied in this study is, for the atmospheric part, the Planet Simulator \citep{Fraedrich2005,Lunkeit2011,Fraedrich2012}. 
In this study, the Planet Simulator is coupled to the Hamburg Large Scale Geostrophic (LSG) ocean model \citep{Maier-Reimer1993}. 
It can also be coupled to other ocean models  \citep[see for example][]{Schmittner2011}, which are, however, not able to simulate as long periods of time as the LSG ocean model. 
The coupled model includes a thermodynamic sea ice model but no model for land ice. 

\subsection{The Planet Simulator}

The Planet Simulator is a spectral model of intermediate complexity, which is freely available under \textit{http://www.mi.uni-hamburg.de/plasim}. 
The dynamical core is based on the Portable University Model of the Atmosphere \citep[PUMA,][]{Fraedrich1998}. 
The primitive equations are solved by the spectral transform method \citep{Eliasen1970,Orszag1970}. 
Unresolved processes are parameterized, which include long-- \citep{Sasamori1968} and short-- \citep{Lacis1974} wave radiation with interactive clouds \citep{Stephens1978,Stephens1984,Slingo1991}. 
Horizontal diffusion according to \citet{Laursen1989} is applied. 
Formulations for boundary layer fluxes of latent and sensible heat and for vertical diffusion follow \citet{Louis1979}, \citet{Louis1982} and \citet{Roeckner1992}. 
Stratiform precipitation is generated in supersaturated states and the Kuo scheme \citep{Kuo1965,Kuo1974} is used for deep moist convection. 
For this study a T21 spectral resolution (approximately \unit{5.6}\degree{} on the corresponding Gaussian grid) with five non-equally spaced vertical $\sigma$-levels is used for the atmosphere. 

Note that the coupled model does not include a model for land ice and ice sheets may not form excluding the effect of mountaineous terrain. 
The snow cover is limited to a height of \unit{5}\meter{} (water equivalent), but the water cycle remains closed, since snow cover which would exceed a depth of \unit{5}\meter{} is fed to the ocean.

\subsection{The LSG ocean model}

The primitive equation ocean model used in this study is the three-dimensional Hamburg Large Scale Geostrophic model (LSG). 
The ocean model is based on the work of \citet{Maier-Reimer1993} and has evolved from an original concept of \citet{Hasselmann1982}. 
The model integrates the momentum equations, including all terms except the nonlinear advection of momentum, by an implicit time integration method that allows a time step of 10~days. 
The free surface is treated prognostically. 
An adaption of the tracer advection scheme by \citet{Farrow1995} for temperature and salinity has been implemented into the model, which uses a predictor--corrector method, with a centered difference scheme for the predictor and a third-order QUICK scheme \citep{Leonard1979} for the corrector stage. 
The QUICK scheme is less diffusive than the standard LSG upstream scheme and less dispersive than the common centered difference scheme. 
Explicit diffusion is necessary to ensure computational stability. 
Depth-dependent horizontal and vertical  diffusivities are employed ranging from \power{10}{7}\centi\squaren\meter\per\second{} at the surface to $5 \times$\power{10}{6}\centi\squaren\meter\per\second{} at the bottom, and from  \unit{0.6}\centi\squaren\meter\per\second{} to \unit{1.3}\centi\squaren\meter\per\second{}, respectively. 
The nonlinear equation of state \citep{UNESCO1981} is used. 
The LSG ocean model is run with 22 vertical $z$-levels (thickness ranges from \unit{50}\meter{} at the top to \unit{1000}\meter{} at the bottom) on a semistaggered grid type ``E'' \citep{Arakawa1977} with $72\times 72$ horizontal grid points, which has a resolution of \unit{3.5}\degree{}. 
For more detailed model description see for example \citet{Prange2003} and \citet{Wenzel2007}. 

\subsection{The coupling}

The coupling is conducted via the surface fluxes of heat, fresh water and momentum. 
No flux correction (or flux adjustment) is applied. 
The interpolation between the atmospheric and the oceanic grid ensures global conservation. 
Atmospheric wind stress and fresh water flux (the sum of precipitation, evaporation and continental runoff) are averaged over the coupling interval (one oceanic time step = 10 days) and then given to the ocean. 
However, it appears that the coupling interval may be too long to keep the oceanic temperature constant during the atmospheric integration. 
Therefore, the uppermost \unit{50}\meter{} of the oceanic column (the thickness of the uppermost layer in the LSG) serve as a mediator for the heat flux: 
During the atmospheric time steps, this part is treated as an oceanic mixed layer and its temperature (heat content) is changed at each time step according to the actual atmospheric heat fluxes (long- and short-wave radiation, as well as sensible and latent heat flux). 
In addition, the temperature is modified by an oceanic heat flux which is computed from the heating/cooling by all oceanic processes during the previous oceanic time step (see below). 
After finishing the atmospheric part, the new temperatures (i.e.\ the new heat content) are provided to the ocean substituting the old ones and the LSG time step is performed. 
The temperature change of the uppermost \unit{50}\meter{} during this time step is translated into an oceanic heat flux which then is used to additionally heat/cool the oceanic mixed layer during the next atmospheric integration (see above). 
Note that if no atmospheric heat flux was present during these next atmospheric steps, the mixed layer temperature (and the oceanic heat content) would consistently obtain the actual LSG value and, thus, energy conservation is ensured.
The evolution of sea ice is computed by the thermodynamic sea ice within the atmospheric part of the integration using the same time step (and grid) as the atmosphere. 
The sea ice thickness is then given to the ocean and provides changes of salinity due to melting or freezing. 

\subsection{Experimental Design}\label{sec:exp}

\begin{figure}[b!]
 \centering
\subfloat[closed \DP{}]{\includegraphics[width=0.35\textwidth]{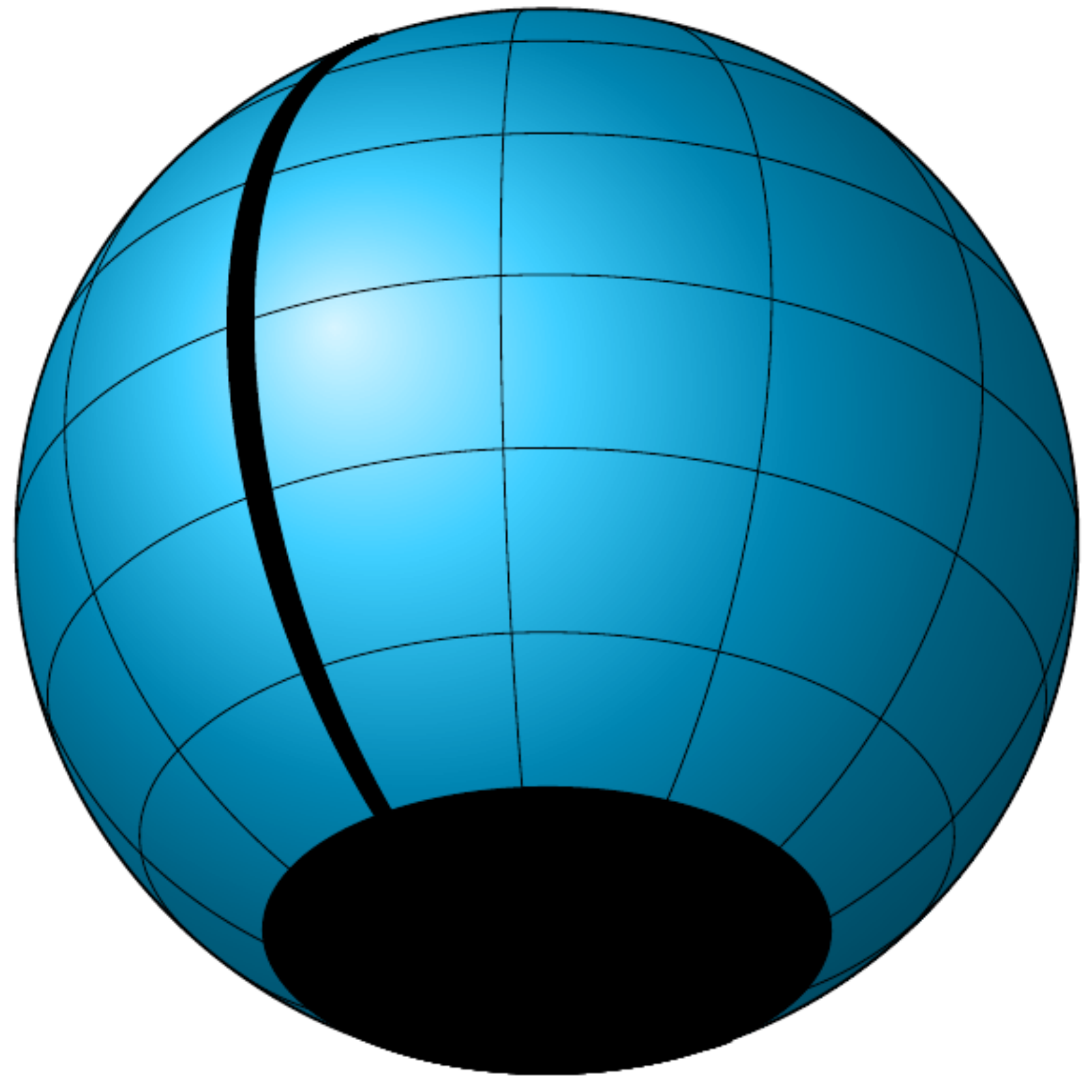}\label{fig:nogap}}\qquad
\subfloat[open \DP{}]{\includegraphics[width=0.35\textwidth]{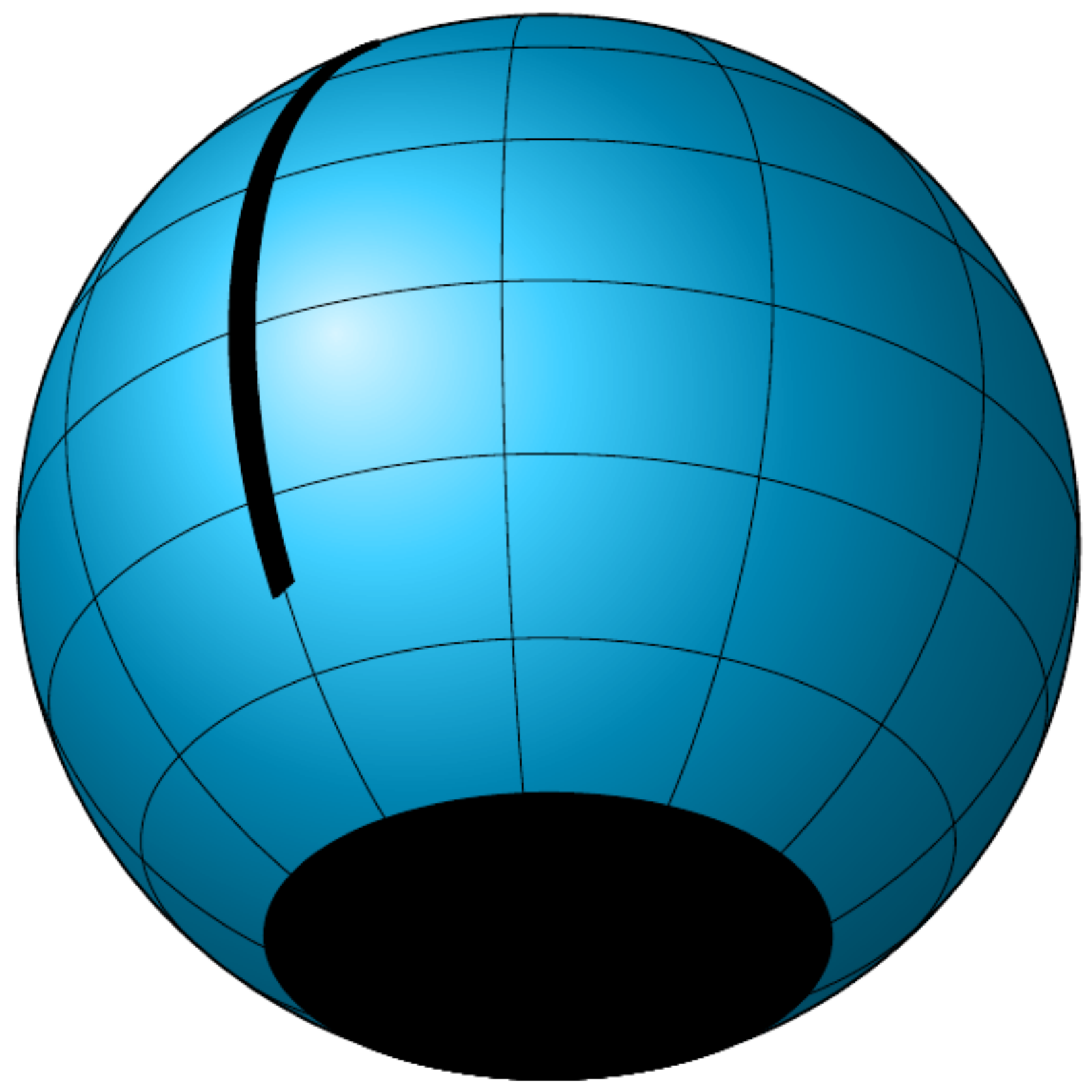}\label{fig:gap}}
\caption{Set-up of aquaplanets with idealized \DP{} (the planet is tilted to highlight the southern ocean and the southern continent)}
\label{fig:DPsetup}
\end{figure}

The simulations are carried out in an idealized land--sea set-up, which expands an aquaplanet model, i.e.\ a planet, where the entire surface of the Earth is covered by one ocean with a flat bottom and without any geometrical constraints. 
The ocean has a uniform depth of \unit{5,500}\meter{} and the solar constant is fixed at \unit{1,365}\watt\per\squaren\meter{}. 
In addition, two topographical features are included, which are displayed in figure~\ref{fig:DPsetup}:
\begin{enumerate}
 \item A circular continent is placed at the south pole (poleward of \unit{60}\degree{}). 
       The continent does not have any orographic features (a uniform height of \unit{0}\meter{} above sea level) nor vegetation cover. 
       However, snow may accumulate on the land surface. 
 \item A meridional barrier (in the ocean only) at \unit{180}\degree{}.
\end{enumerate}

Three variations in the boundary conditions are chosen:

\begin{itemize}
 \item \textbf{Opening of an ocean gateway:} The meridional barrier runs from the north pole down to the continent in the case of a closed gateway (figure~\ref{fig:nogap}). 
For an opened gateway (figure~\ref{fig:gap}), the barrier runs down to \unit{30}\degree{}S and thus leaves a gap in the southern ocean. 
Since the passage resembles a very idealized case of a Drake Passage, it will be called \DP{} (in italic) in the following. 

 \item \textbf{Change in atmospheric \CO{}:} The change in atmospheric \CO{} is represented by comparing simulations with present-day \CO{} (\unit{360}\ppm{}) and cases with \CO{} values of \unit{1000}\ppm{}. 

 \item \textbf{Change in orbital parameters:} The change in the orbital parameters is represented by simulations with and without seasonality. 
The cases with seasons have present-day orbital parameters (eccentricity of 0.0167 and obliquity of 23.441). 
In the cases without seasons perpetual equinoctial conditions are created by setting the obliquity and the orbital eccentricity to zero. 
These changes in the orbital parameters cause great differences in the insolation forcing, which are especially prominent at high latitudes. 
To futher analyze the impact of the orbital parameters, additional simulations with obliquities of 11.5 and 22.5 are analyzed. 

\end{itemize}

\begin{figure}[tbh]
\includegraphics[width=\textwidth]{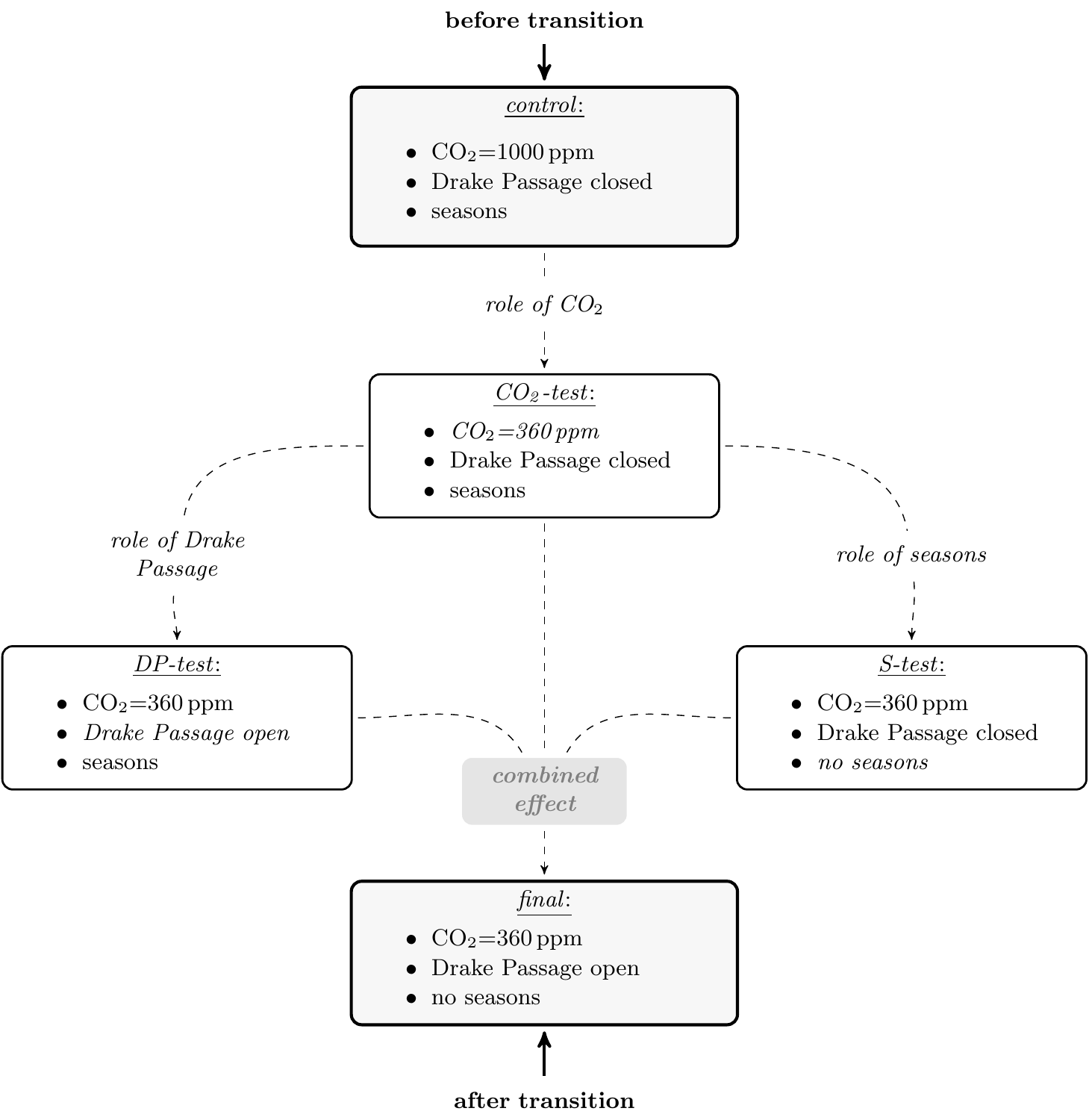}
\caption{Overview of experimental design}
\label{fig:expoverview}
\end{figure}

An overview of the experiments is given in figure~\ref{fig:expoverview}. 
A (fictional) climate transition is simulated that resembles the case of the Antarctic glaciation:
The \contr{} experiment is the state of the climate system before the transition: high \CO{} levels, a closed \DP{}, and orbital parameters that cause seasonality. 
After the transition the climate has shifted to the so-called \final{} state: lower \CO{} levels, an open \DP{}, and orbital parameters that suppress seasonality. 

A comparison between the \final{} and \contr{} case will be presented. 
Furthermore, sensitivity experiments are analyzed, where the different effects are studied separately:
\begin{enumerate}
 \item change in atmospheric \CO{} concentrations only: \COtest{}
 \item\label{item:DP} role of the \DP{} only (with a lower \CO{} level): \DPtest{}
 \item\label{item:S} role of the orbital parameters only (with a lower \CO{} level): \Stest{} (as well as \Stestelf{} and \Stestzz{})
 \item linear combination of the effects of decreasing \CO{}, opening the \DP{}, and changing the orbital parameters. 
\end{enumerate}

\afterpage{\clearpage}

The changes in the three forcings are not realized with a transient simulation but with five individual experiments, each with constant parameters and boundary conditions. 
Each simulation starts from steady state and is integrated for 20,000~years to ensure that it has reached equilibrium. 
The spin-up period of the ocean is very long and to still obtain a robust average, the mean climate state is taken over the last 1,000 years (years 19,001 to 20,000). 

\section{Mean climate states under different forcings}
\label{sec:results}

\subsection{Before and after the transition}

Two simulations are integrated out to equilibrium to represent the states before and after the climate transition. 
The \contr{} forcing is characterized by high \CO{} values, a closed \DP{} that inhibits circumpolar flow around the southern polar continent, and orbital parameters that allow a yearly cycle with pronounced seasonality. 
During the transition atmospheric \CO{} values decline dramatically, the \DP{} opens up, and the orbital parameters change suppressing the seasonal cycle. 
The climate state after this transition is represented in the \final{} simulation. 
Here, the mean climate states of \contr{} and \final{} are introduced and compared. 

\begin{figure}[bt]
 \centering

\includegraphics[width=0.65\textwidth]{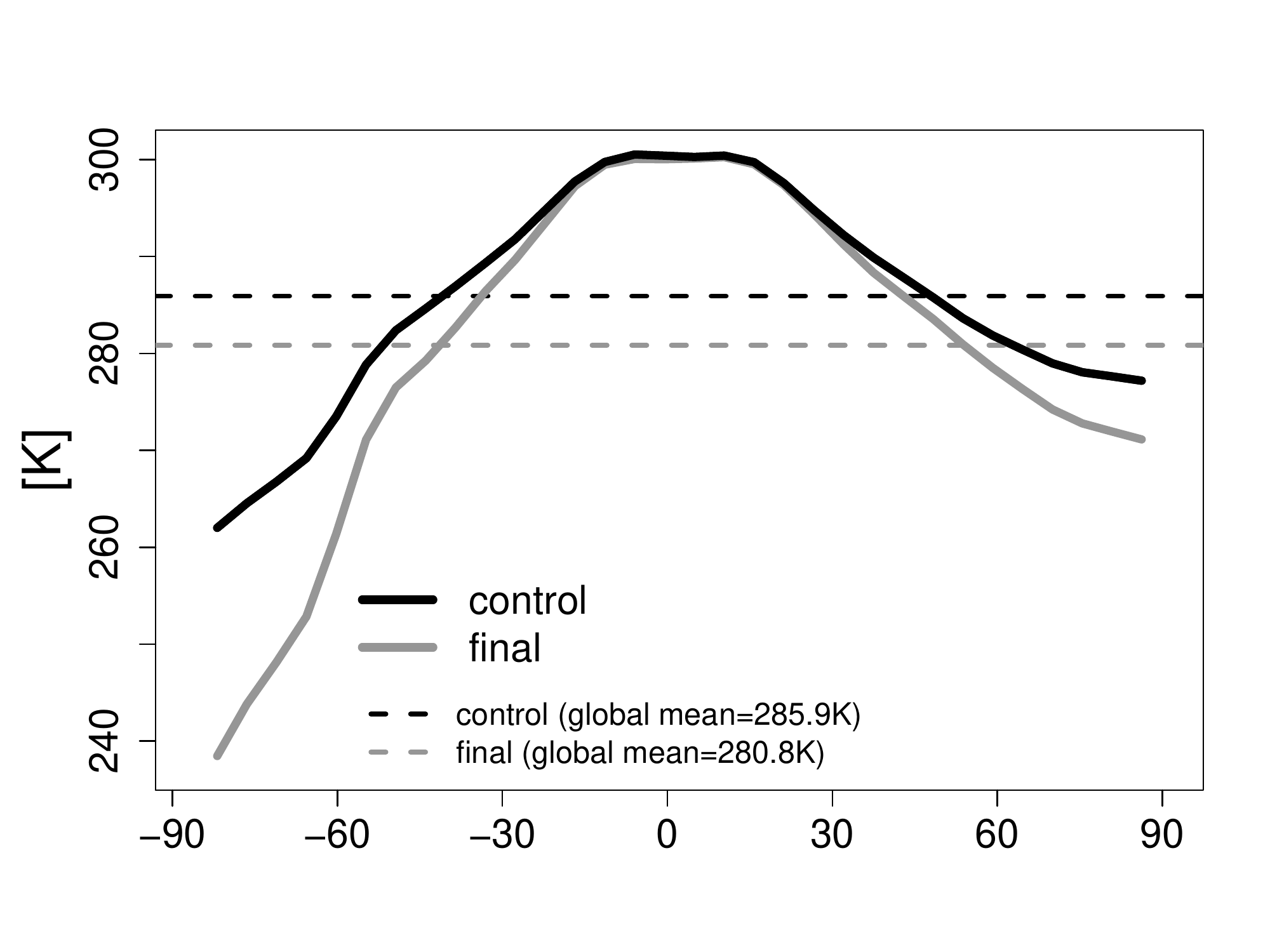}

\caption{Zonal and global mean surface air temperature for \contr{} and \final{}}
\label{fig:SATzm}
\end{figure}

\begin{figure}[th!]
 \centering

\subfloat[atmospheric temperature (in \kelvin{})]{\includegraphics[width=0.45\textwidth]{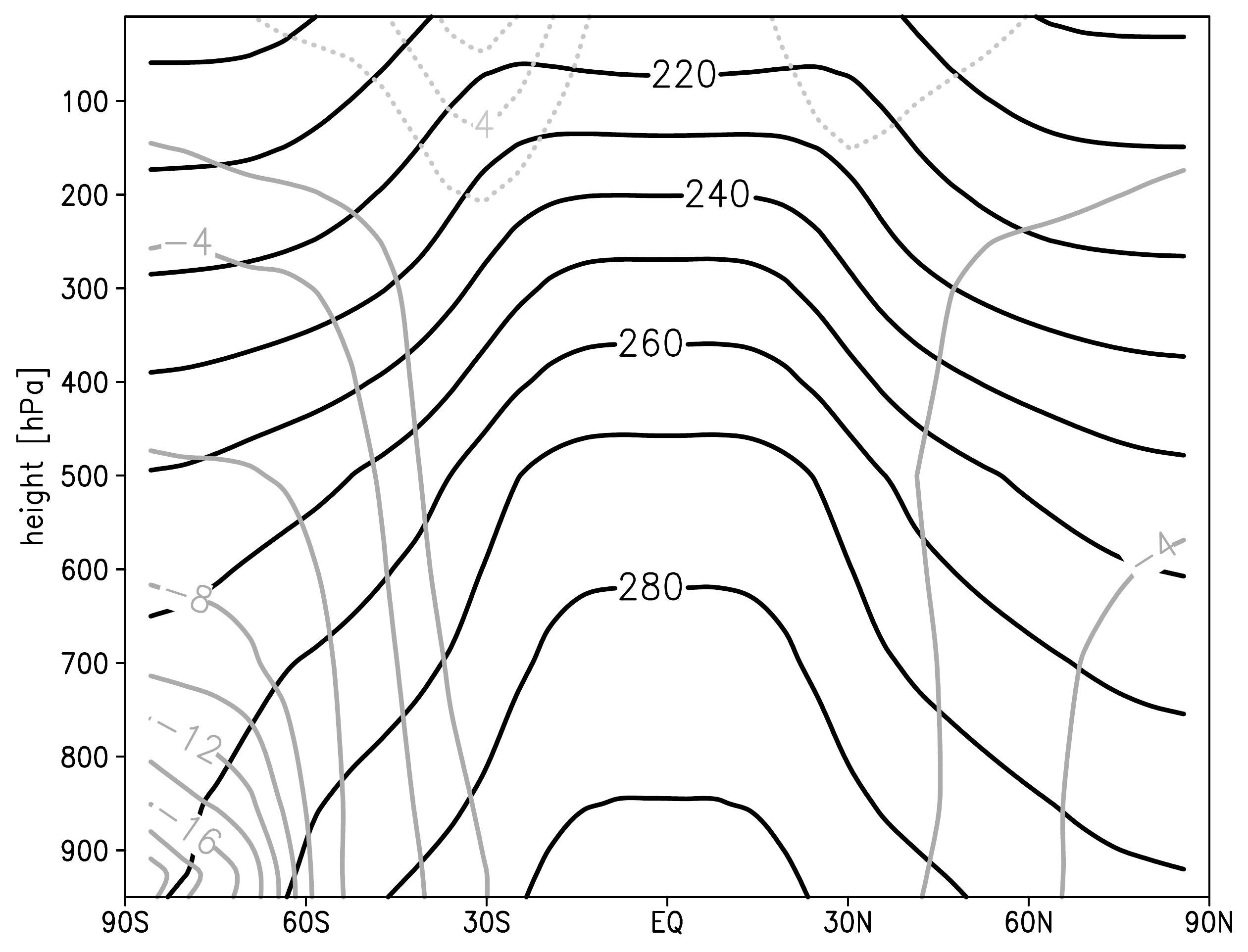}\label{fig:atmT}}\quad
\subfloat[atmospheric stream function (in \power{10}{9}\kilo\gram\per\second{})]{\includegraphics[width=0.45\textwidth]{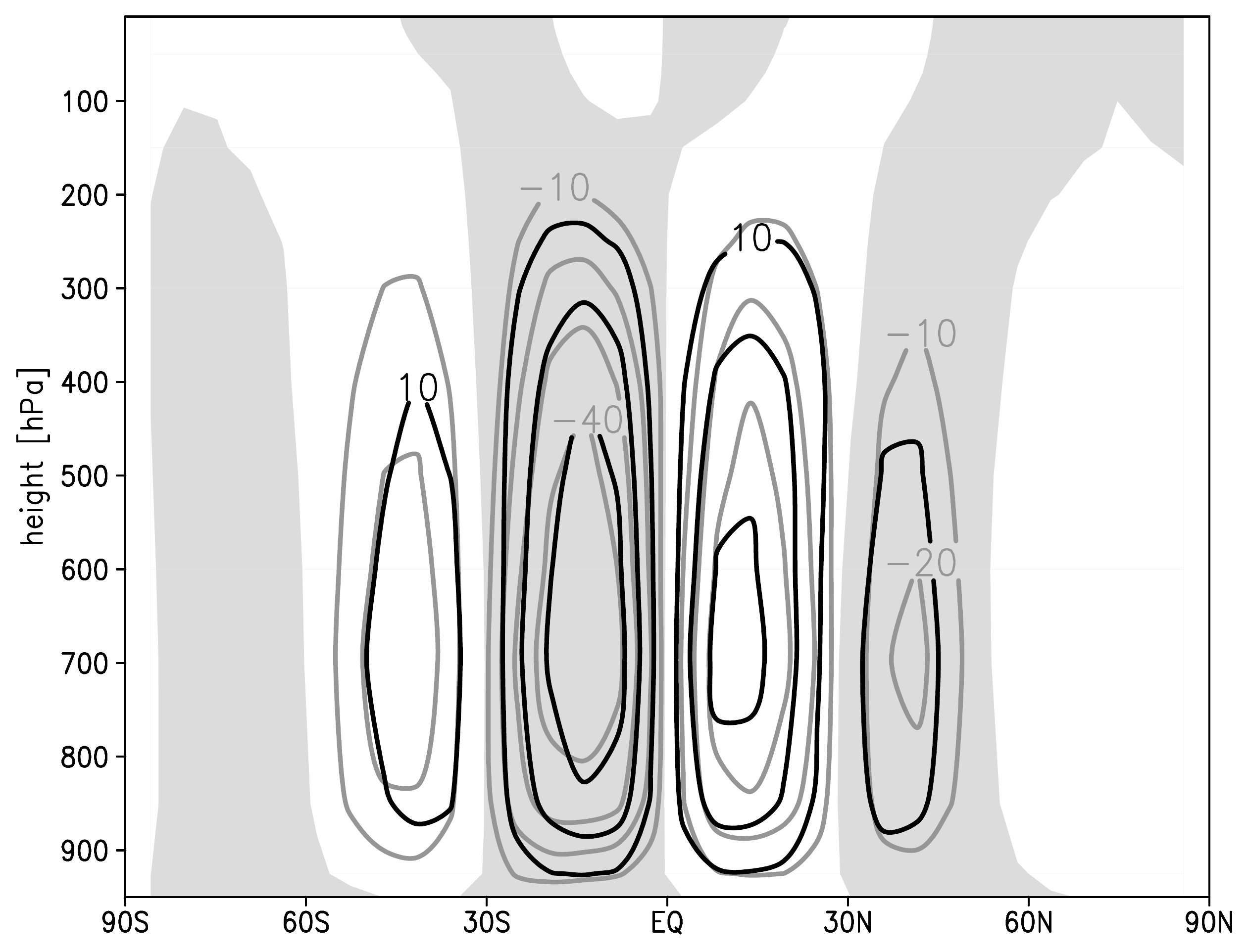}\label{fig:atmPSI}}

\subfloat[ocean temperature (in \kelvin{}, up to \unit{2.2}\kilo\meter{} depth)]{\includegraphics[width=0.45\textwidth]{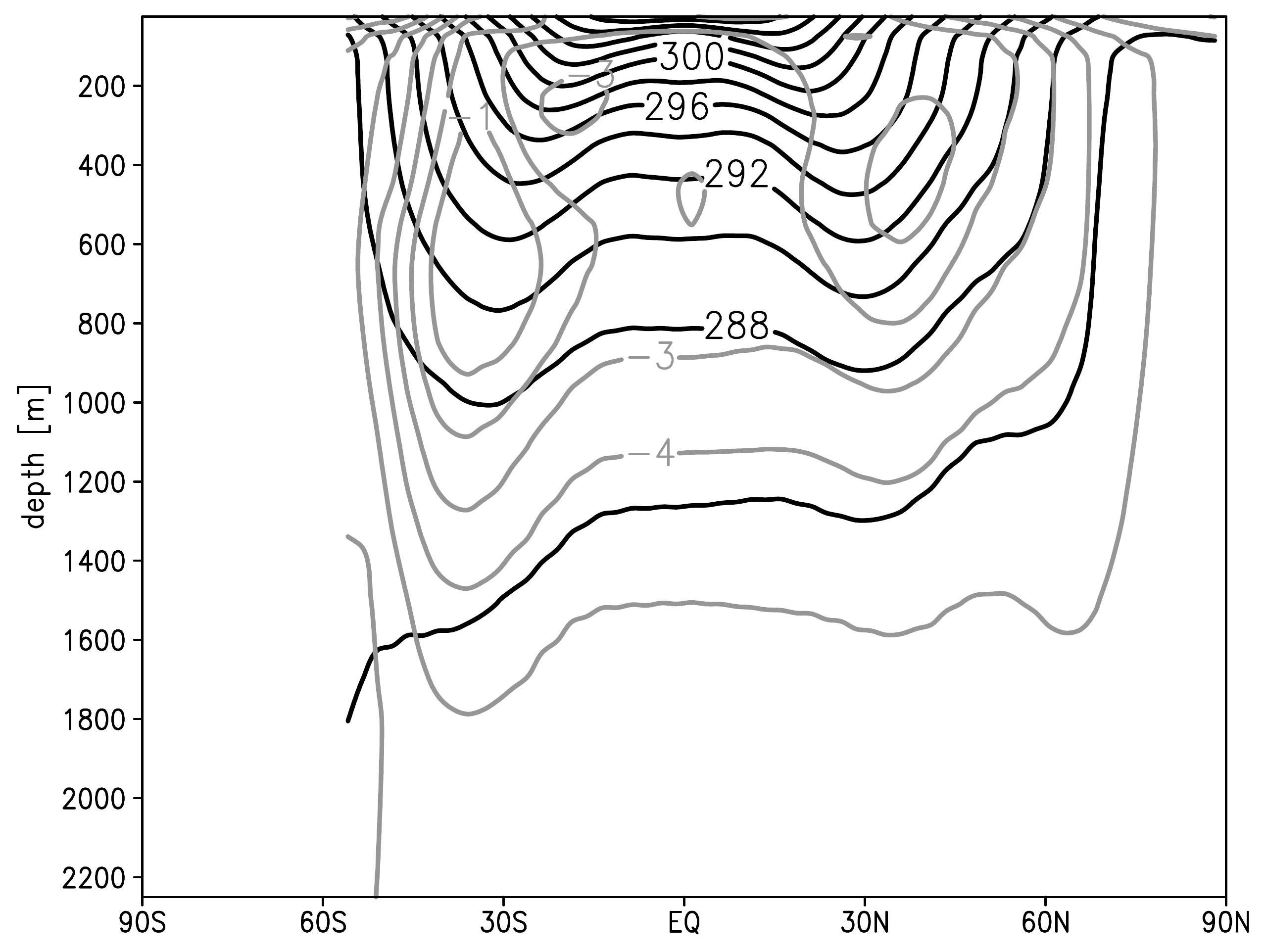}\label{fig:ocT}}\quad
\subfloat[ocean MOC (in \sv{})]{\includegraphics[width=0.45\textwidth]{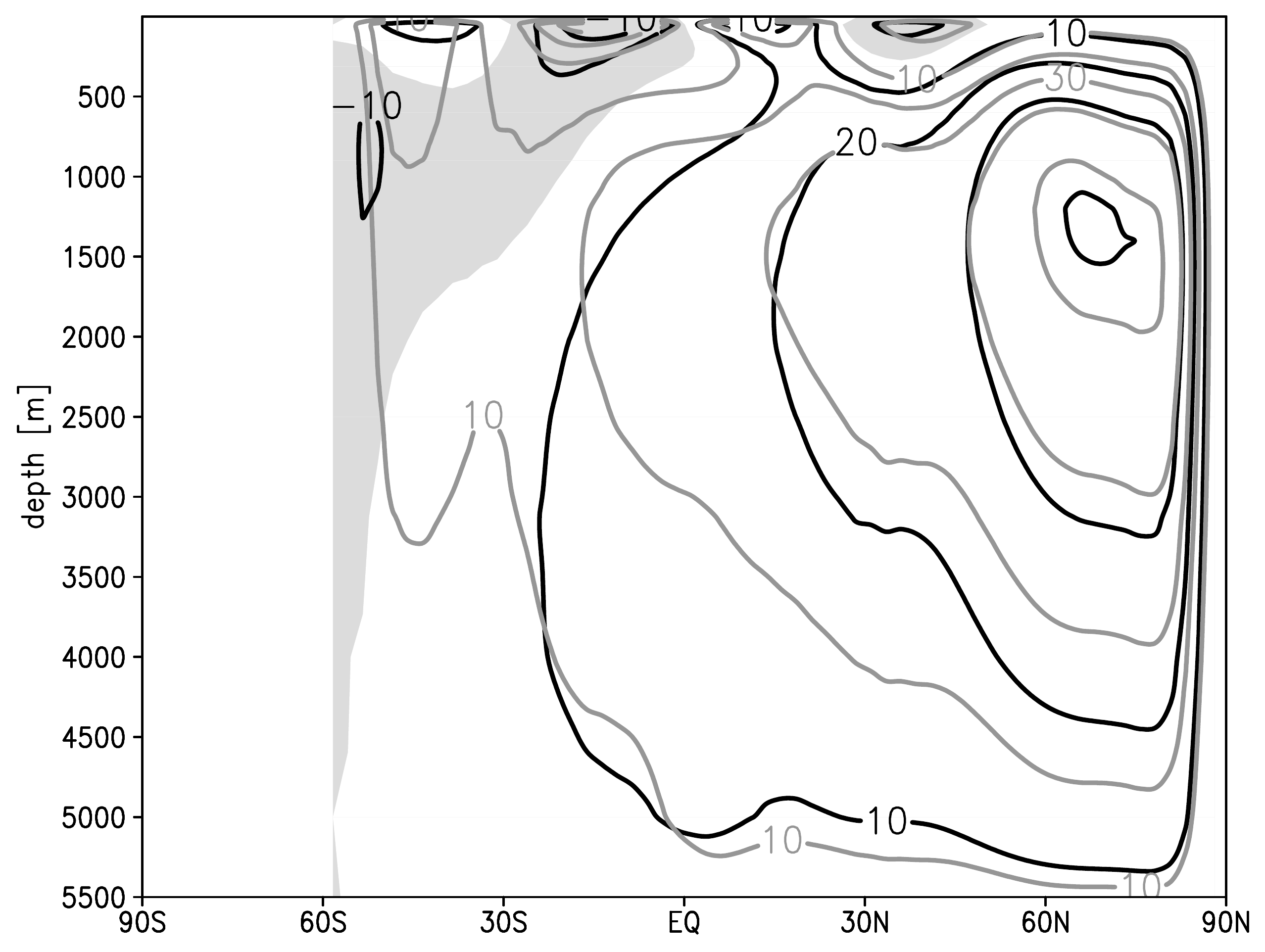}\label{fig:ocPSI}}

\caption{Zonal mean temperature in (a) atmosphere and (c) ocean (\contr{} is represented by black and differences \final{}--\contr{} by gray lines; positive deviations are denoted by dotted lines) as well as (b) atmospheric stream function and (d) MOC (shaded areas denote negative stream functions in the \contr{} climate; black lines represent the \contr{} and gray lines the \final{} climate)}
\label{fig:T-PSI}
\end{figure}

\begin{figure}[htb!]
 \centering

\includegraphics[width=0.65\textwidth]{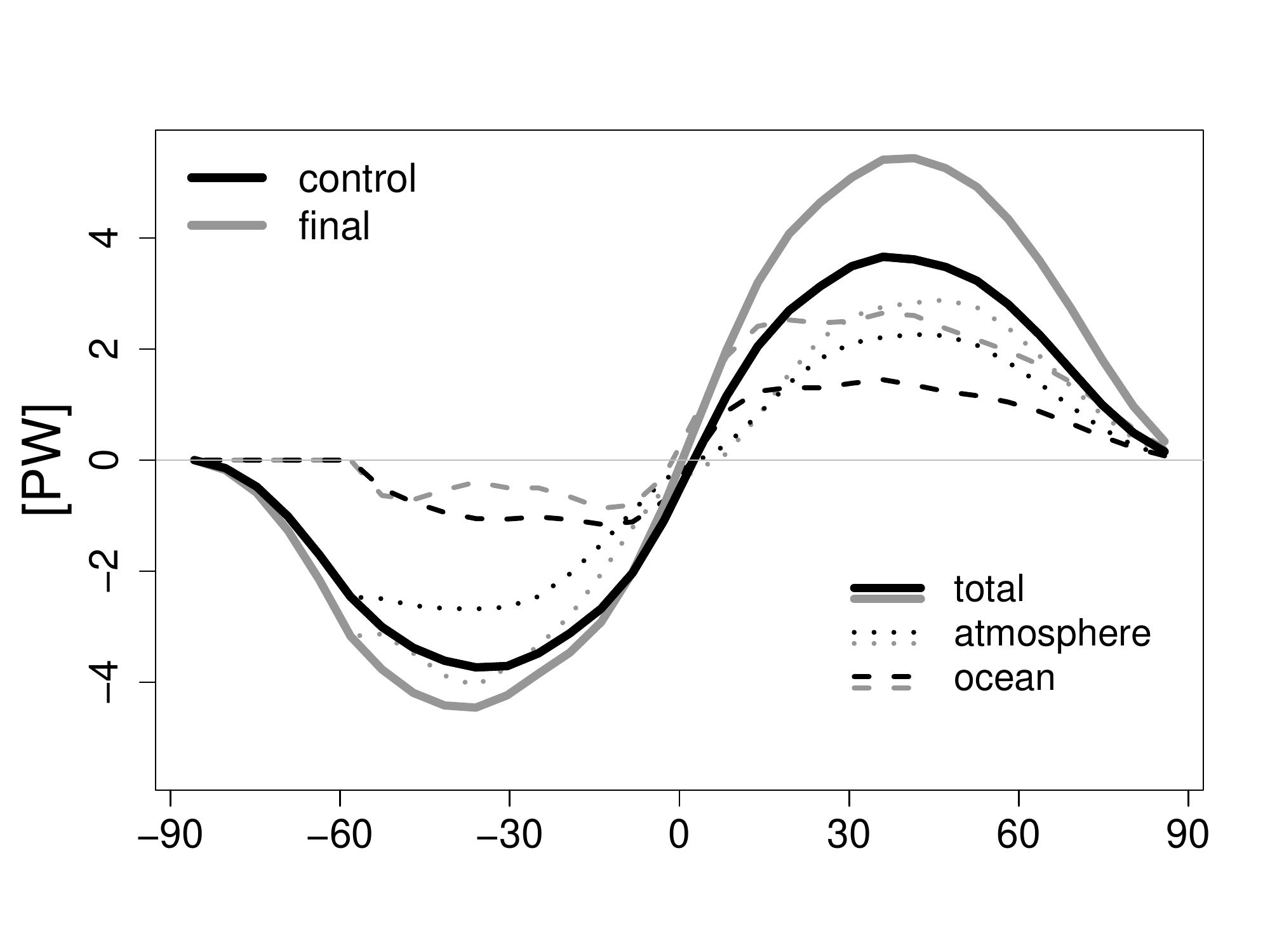}

\caption{Vertically integrated and zonally averaged meridional energy transport for \contr{} and \final{}: total transport and the partition into atmospheric and oceanic transport}
\label{fig:transport}
\end{figure}

Features of the zonal mean climate of the two cases (\contr{} and \final{}) are displayed in figures~\ref{fig:SATzm}, \ref{fig:T-PSI}, and \ref{fig:transport}.
The surface air temperature (SAT) cools in the global mean by approximately \unit{5}\kelvin{} from \contr{} to \final{}. 
However, as seen in figure~\ref{fig:SATzm}, the SAT does not decline equally at all latitudes. 
In the tropics and subtropics, both climate states show almost the same meridional temperature profile, but at mid- and higher latitudes, the temperatures in \final{} are already significantly lower than in \contr{}. 
In the northern hemisphere, the differences amount up to \unit{5}\kelvin{}. 
The greatest discrepancies in the meridional temperature profiles can be observed at the south pole over the continent, where a cooling by \unit{10--20}\kelvin{} takes place. 

In the tropics, the zonal mean atmospheric temperature distribution at all heights (figure~\ref{fig:atmT}) hardly differs between the two climate states before and after the transition. 
Even though \CO{} is drastically decreased, the temperatures at low latitudes hardly cool. 
This is due to compensation effects of the opening of the \DP{} and the change in the orbital parameters (see sections~\ref{sec:CO2}, \ref{sec:DP}, and \ref{sec:S}). 
Poleward of approximately \unit{30}\degree{}N/S, the temperature distributions start to deviate significantly and neither the temperature profiles of \contr{} nor the temperature deviations to \final{} are symmetric about the equator anymore (especially in the lower atmosphere). 
The northern mid- and high latitudes of \contr{} are warmer than their counterparts in the southern hemisphere, where especially the temperatures close to the surface become very cold. 
Furthermore, the \final{} climate is much colder than the \contr{} climate at mid- and high latitudes, especially over the continent at the south pole where differences exceed \unit{10}\kelvin{} up to the middle troposphere. 
Only in the upper subtropical atmosphere, the temperature increases in the \final{} climate. 

In the ocean the temperature decreases at all latitudes and at all depths (figure~\ref{fig:ocT}, only depths up to \unit{2200}\meter{} are displayed) between \contr{} and \final{}. 
Contrasts are especially high (over \unit{5}\kelvin{}) in the area of the closed/open \DP{} and in the deep ocean. 

Sea ice does not form in neither northern nor southern polar waters. 
In \contr{} the sea surface temperatures (SSTs) are much too warm even at the poles (above \unit{278}\kelvin{} all year long). 
In \final{} the SSTs at high latitudes are much colder but still not below freezing. 

The annual mean atmospheric stream function (figure~\ref{fig:atmPSI}) shows a direct Hadley circulation in the lower and an indirect Ferrel cell in the mid-latitudes. 
The strength and structure of the Hadley cells are strongly connected to the equator-to-pole temperature gradient, which can be explained, for example, by the model of \citet{Held1980}.
The Hadley cells have a maximum strength of $30-40\times$\power{10}{9}\kilo\gram\per\second{} (the atmospheric stream function in \power{10}{9}\kilo\gram\per\second{} represents numerically about the same mass transport as an oceanic volume transport in Sverdrups; \citealt{Czaja2006}), which is comparable to present-day annual mean observations  \citep[for example see][]{Peixoto1992}. 
Since the Hadley cell is connected to the meridional temperature gradient, in each climate state the cell in the southern hemisphere is stronger than in the northern hemisphere. 
The southern Hadley circulation in the \final{} climate is the strongest with a maximum of $40\times$\power{10}{9}\kilo\gram\per\second{}, because of the great temperature difference between the very cold southern polar continent and the warm tropics. 
\citet{Lu2007} analyze the Hadley circulation under greenhouse gas forcing and also observe a weakening under increased greenhouse gases. 
The Ferrel cells have about half the strength of the Hadley cells (maxima of $10\times$\power{10}{9}\kilo\gram\per\second{} in the \contr{} simulation and $20\times$\power{10}{9}\kilo\gram\per\second{} in the \final{} climate after the transition). 

The meridional overturning circulation in the ocean (MOC), which is displayed in figure~\ref{fig:ocPSI}, is very much affected by the opening of the \DP{}.  
In a pure aquaplanet (no orographic barriers at all), a tropical and an extra-tropical circulation cell develops on each side of the equator \citep[for a discussion of pure aquaplanets see][]{Smith2006, Marshall2007,Hertwig2015}. 
In this case with a meridional barrier in the ocean, a large cross-equatorial (positive) cell develops in the northern hemisphere (in each simulation), which reaches into the subtropics of the southern hemisphere. 
Only close to the surface there is a small negative cell in the northern hemispheric mid-latitudes. 
In the \contr{} climate, when the \DP{} is closed, a (negative) cell develops in the southern hemisphere, which is the counterpart to the huge northern cell, even though it is much smaller and weaker (and of course bounded by the southern continent). 
In the case where the \DP{} opens up (the \final{} climate), this southern cell is confined to the very surface and a stronger and larger mid-latitude cell can be observed, which is connected to the northern hemispheric tropical cell. 
Compared to other idealized studies \citep[e.g.][]{Marotzke1991,Toggweiler2000,Smith2006,Enderton2009}, the MOC in \contr{} and in \final{} show similar patterns and magnitudes but slightly higher maxima. 

\begin{figure}[th!]
 \centering

\subfloat[SAT]{\includegraphics[width=0.65\textwidth]{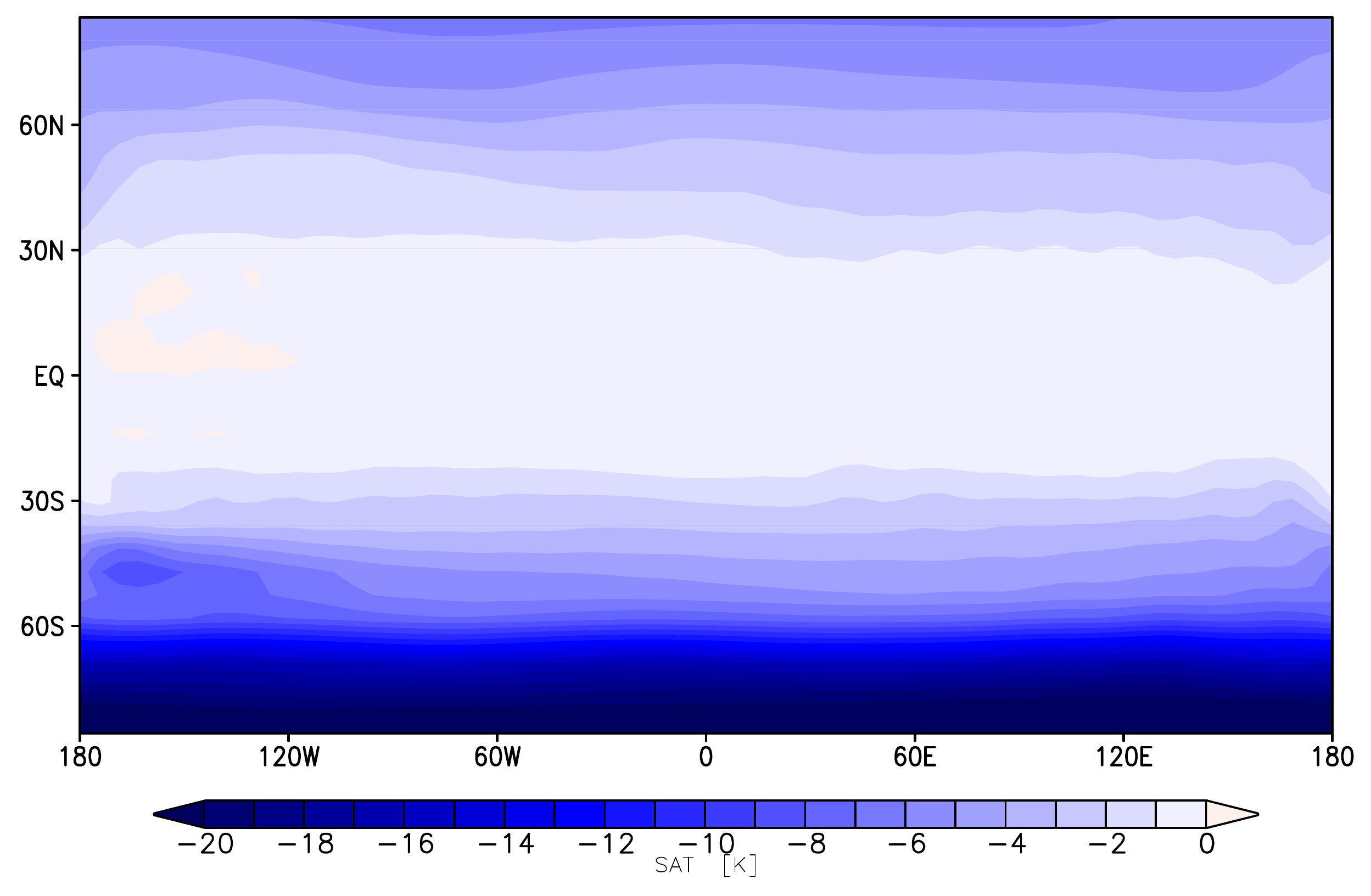}\label{fig:final-contr_SATdiff}}

\subfloat[snow]{\includegraphics[width=0.75\textwidth]{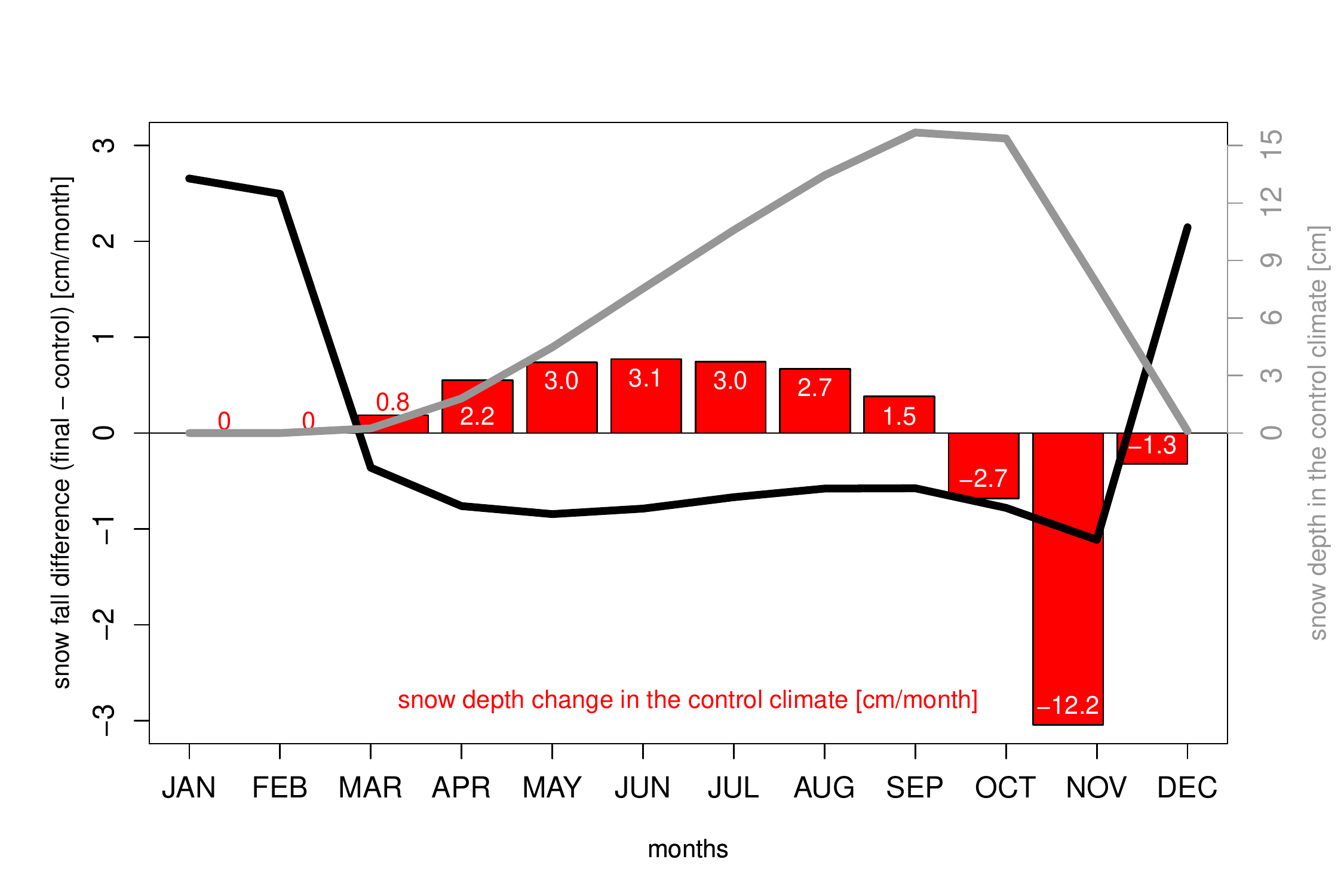}\label{fig:final-contr_snow}}

\caption{Differences between the \final{} and \contr{} climate states: (a) annual mean global surface air temperature differences and (b) monthly mean snowfall, snow depth, and snow depth change averaged over the continental area (all snow values are in liquid water equivalent units)}
\label{fig:final-contr_SATsnow}
\end{figure}

The zonal mean vertically integrated transport of energy is displayed in figure~\ref{fig:transport}, as well as the partition of the transport between atmosphere and ocean. 
\citet{Bjerknes1964} and \citet{Stone1978} explain that the sum of the oceanic and atmospheric meridional energy transports remains mostly constant throughout changes imposed on the climate system. 
However, \citet{Stone1978} found that the controlling factors on the strength and form of the transport are size of the planet, rotation rate, axis tilt, the solar constant, and the mean hemispheric albedo. 
Various other studies \citep[e.g.][]{Manabe1975,Clement1999,Winton2003} support this hypothesis. 
While the solar constant, the size and rotation rate of the planet do not change, there are differences in the axis tilt of the planet and in the mean hemispheric albedo. 
In the \contr{} climate the axis of the Earth is tilted by approximately \unit{23.5}\degree{}, but in \final{} the rotation axis is perpendicular to the orbital axis of the Earth (tilt of \unit{0}\degree{}). 
Additionally, the zonal mean albedo above the continent changes (from about 0.4 to 0.6) because of differences in snow cover (see figure~\ref{fig:final-contr_snow}). 
Therefore, after Stone's (\citeyear{Stone1978}) concept, it is not surprising that in the climates of \contr{} and \final{} very different amounts of total energy are transported poleward. 
In the \contr{} climate, the total heat transport peaks at almost \unit{4}\peta\watt{} between \unit{30}\degree{} and \unit{35}\degree{} in each hemisphere, but with a slightly stronger transport in the south. 
After the transition, the total zonal mean meridional energy transport has increased dramatically. 
The maxima in the \final{} climate exceed \unit{5}\peta\watt{} and are located around \unit{40}\degree{}N/S, with higher values in the northern hemisphere. 

In the northern hemisphere, the ocean heat transport (OHT) is very large and almost as high as the atmospheric energy transport. 
In contrast, in the southern hemisphere, the atmospheric heat transport is significantly larger than the oceanic part. 
The \contr{} climate has a relatively constant OHT of about \unit{1}\peta\watt{} in the lower and mid-latitudes in both hemispheres. 
The atmospheric transport is higher, especially in the mid-latitudes over the continent, and even more in the southern polar latitudes, where no OHT can occur. 
In the northern tropics and subtropics, the OHT in the \final{} climate reaches \unit{2}\peta\watt{} and even exceeds the atmospheric transport, which peaks at higher latitudes.  
In the southern hemisphere, the OHT is significantly reduced compared to the northern hemisphere. 
In the area where the continent is located (poleward of \unit{60}\degree{}S), the atmosphere conducts the entire heat transport alone. 
Furthermore, especially in the case where the \DP{} is opened and a strong circumpolar current develops, the OHT is very small and the dominant contribution of the poleward heat transport has to be achieved by the atmosphere. 

\begin{figure}[th!]
 \centering

\subfloat[\contr{}]{\includegraphics[width=0.65\textwidth]{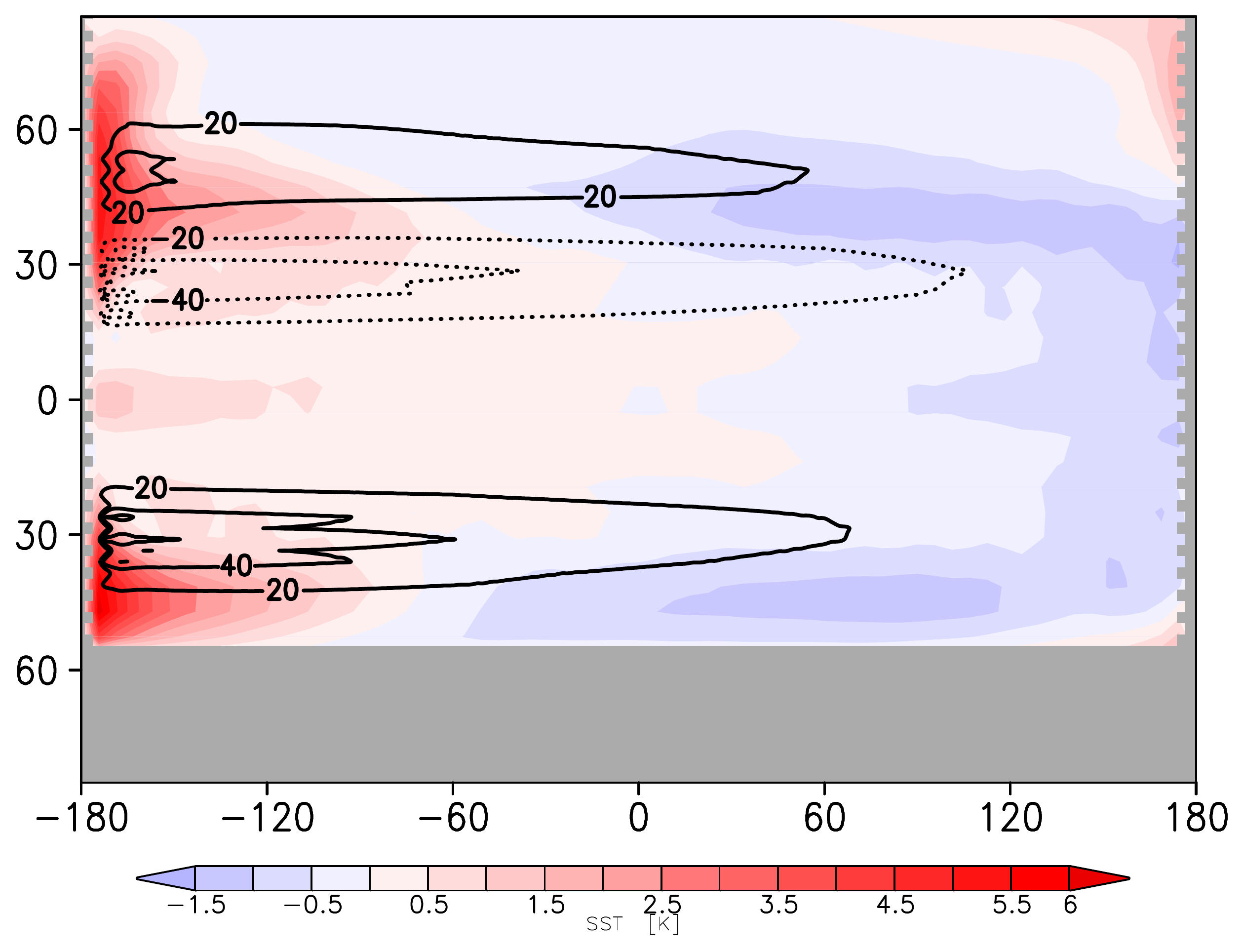}\label{fig:nogapCO2_flow-sst}}

\subfloat[\final{}]{\includegraphics[width=0.65\textwidth]{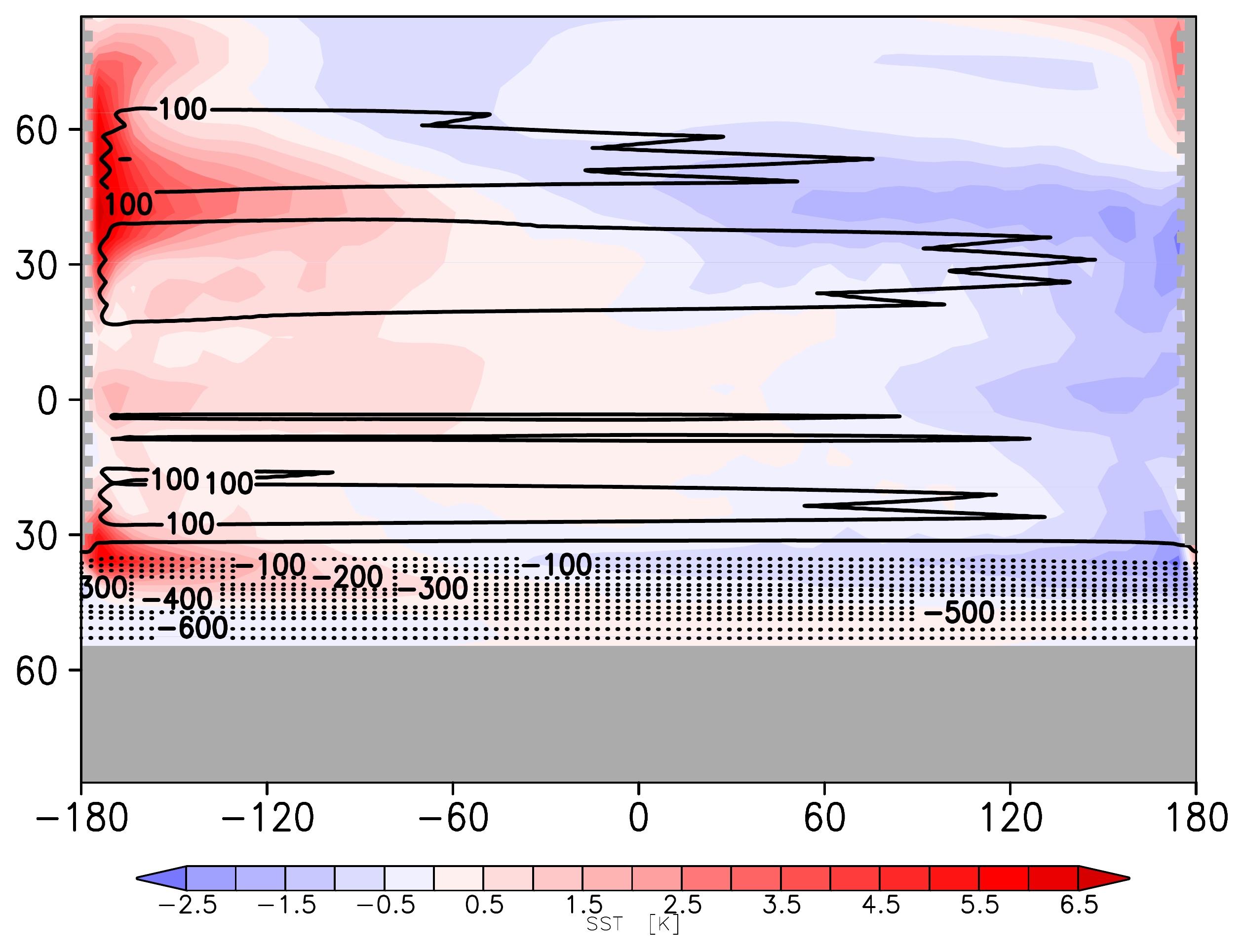}\label{fig:gap1_flow-sst}}

\caption{Deviation of the sea surface temperature from its zonal average and barotropic horizontal stream function (spacing of the stream lines is \unit{20}\sv{} in the \contr{} and \unit{50}\sv{} in the \final{} case)}
\label{fig:flow-sst}
\end{figure}

For both \contr{} and \final{} climates, figure~\ref{fig:flow-sst} shows the horizontal barotropic stream function and the deviation of the sea surface temperature (SST) from its zonal average. 
Northward of \unit{30}\degree{}S, both climates reveal approximately the same SST deviation patterns, even though in the \final{} state SST contrasts are more pronounced. 
Warmer surface waters are located at the western boundary of the ocean basin, while the eastern boundary is much colder. 
The only exception is the northern polar ocean, which is warm close to both sides of the boundary and colder in the middle of the ocean basin. 
In the mid-latitudes the warm/cold contrast between west and east is especially strong. 

The barotropic stream function (depth integrated flow) is directly related to the pattern of surface wind stress. 
In the northern hemisphere, both \contr{} and \final{} show a subtropical and a subpolar gyre with greater magnitudes in the \final{} climate: up to \unit{100}\sv{} compared to approximately \unit{40}\sv{} in \contr{} (note the different contour line spacings for \contr{} and \final{}). 
These values are similar to the ones found by \citet{Enderton2009}, who observe maximum gyres between \unit{80}\sv{} and \unit{100}\sv{} in their \DP{} simulation. 
\citet{Enderton2009} and \citet{Smith2006} also observed a strengthening of the gyres when opening the \DP{} in their respective set-ups. 
The intensification of the gyres in \final{} goes along with a stronger MOC (see figure~\ref{fig:ocPSI}). 

The subtropical gyre can also be found in the southern hemisphere of both \contr{} and \final{}. 
The greatest deviations between \contr{} and \final{} occur poleward of the southern mid-latitudes: 
In the \contr{} case (with a closed \DP{}, figure~\ref{fig:nogapCO2_flow-sst}) the SST distribution is almost a mirror of the northern hemisphere (equatorward of the continent). 
In contrast, in the \final{} climate, when the \DP{} is opened (figure~\ref{fig:gap1_flow-sst}), hardly any zonal temperature deviations exist poleward of \unit{30}\degree{}S. 
This can be explained by the different circulation patterns: 
When the \DP{} in the southern ocean opens up a very strong circumpolar current develops, which is missing in the set-up with the closed barrier. 
This depth integrated circumpolar flow is very strong (more than \unit{500}\sv{} are transported), which is consistent with other idealized model studies comprising a flat ocean bottom (see for example \citealt{Smith2006} or \citealt{Enderton2009}). 

Figure~\ref{fig:final-contr_SATdiff} shows the differences in surface air temperature (SAT) between the \contr{} and the \final{} climate. 
During the transition the climate becomes colder at all latitudes, even though the cooling in the tropics is only very weak (in the range of \unit{1--2}\kelvin{} and even a minimal warming at the western boundary of the ocean basin). 
Mid- and especially high latitudes of the northern hemisphere experience a strong temperature reduction of up to \unit{6}\kelvin{}. 
However, the cooling in the southern hemisphere is even larger. 
Poleward of \unit{30}\degree{}S the SAT cools significantly, especially at the western boundary of the ocean basin. 
The main cooling occurs over the continental area, where the temperature drops between \unit{10}\kelvin{} and \unit{20}\kelvin{}. 

The differences in snowfall averaged over the continental area are displayed in figure~\ref{fig:final-contr_snow}. 
A strong seasonal cycle can be observed:
In the austral summer months, between December and February, there is significantly more snowfall in the \final{} climate (almost \unit{3}\centi\meter{} per month more). 
During the rest of the year, the snowfall is reduced compared to the \contr{} climate, however, the reduction is \unit{1}\centi\meter{} per month at most. 
Accumulated over the year, there is still an increase in the snowfall of \unit{0.8}\centi\meter\per\yr{}. 

Not surprisingly, the snow depth is much higher in the \final{} than in the \contr{} climate. 
The snow depth (for the \contr{} climate) is also displayed in figure~\ref{fig:final-contr_snow} (differences are not shown since the two climate states have different orders of magnitudes of snow depths). 
In \contr{} there is a strong seasonal cycle. 
Maximum snow depths can be observed in the late winter/early spring months (September and October) with approximately \unit{14--15}\centi\meter{} (all values are liquid water equivalent depths). 
In the austral summer months (between December and March) there is no snow cover on the southern continent. 
In the \final{} climate, there is constant snow accumulation and the snow depth would grow continuously if it was not limited by the model (see the model description in section~\ref{sec:model}). 
Thus, the snow depth in \final{} is constant at \unit{5}\meter{}. 

Additionally, the snow depth change (accumulation or melting rate in \centi\meter\per{}month) is shown in figure~\ref{fig:final-contr_snow} (for the \contr{} climate only; for the \final{} climate this parameter is zero since snow accumulation over \unit{5}\meter{} is balanced by artificial melting). 
The snow depth change illustrates how important the melting period in the austal spring/early summer time is. 
During most of the year, snow accumulates on the southern continent. 
However, just when the show depth is highest, the temperatures start to increase and the snow depth decreases (with a maximum rate over \unit{12}\centi\meter{} per month in November). 
When the temperatures are highest, in December and January, there is already no snow cover left. 

In summary, the experiments show that the differences between \contr{} and \final{} are pronounced and a transition from the one climate state to the other leads to much colder temperatures and higher snow covers, especially over the southern polar continent. 
However, the two simulations, \contr{} and \final{}, differ in three parameters: the atmospheric \CO{} concentration, the closed/open \DP{}, and the orbital parameters. 
To analyze the relative importance of each parameter and to explore which is the main forcing mechanism for the cooling of the southern continent, three further sensitivity experiments are discussed. 

\subsection{The role of atmospheric \CO{}}\label{sec:CO2}

First of all, we analyze the role of the atmospheric \CO{} decrease during the transition. 
A simulation with exactly the same set of parameters as \contr{}, but with lower \CO{} values (\unit{360}\ppm{}, called \COtest{}) is compared to the \contr{} climate (\unit{1000}\ppm{}). 

\begin{figure}[tb]
 \centering

\includegraphics[width=0.65\textwidth]{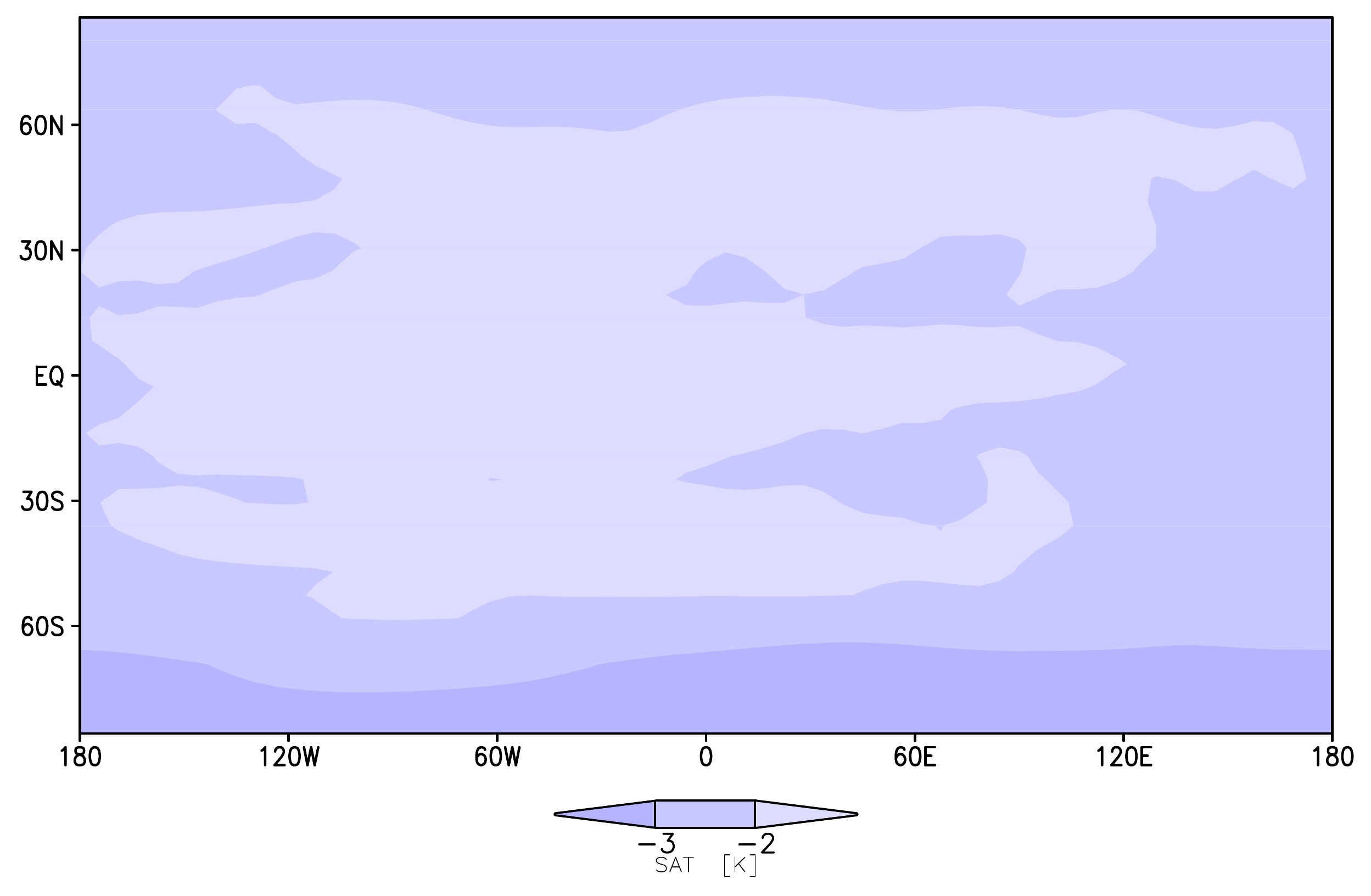}

\caption{Surface air temperature difference between the \COtest{} and the \contr{} climate state}
\label{fig:CO2test-contr_SATdiff}
\end{figure}

Figure~\ref{fig:CO2test-contr_SATdiff} shows the surface air temperature difference between the \COtest{} and \contr{} climate states. 
When atmospheric \CO{} is reduced, the temperature becomes colder at all latitudes. 
The cooling is globally relatively constant with a temperature reduction of \unit{2--3}\kelvin{}.  
Even though the reduction shows the largest values over the continent, the magnitude is not sufficient to explain the massive temperature decrease between the \contr{} and \final{} climate. 
Therefore, it can be concluded that decreasing atmospheric \CO{} does not force the cooling of the southern continent alone. 

Annual mean snow parameters as well as surface albedo and temperature for all sensitivity experiments are shown in table~\ref{tab:snow}. 
There is slightly more snow fall in \COtest{} (compared to \contr{}), but mostly the snow and surface parameters only change slightly when \CO{} is decreased.

\subsection{The role of the ocean gateway}\label{sec:DP}

The cooling effect of atmospheric \CO{} decrease is relatively small compared to the overall cooling during the transition (only about \unit{10}\%{}). 
Therefore, other mechanisms must play a more crucial role. 
The role of the opening of the \DP{}, as well as the role of the orbital parameters (see section~\ref{sec:S}), are analyzed separately. 
Here, the \COtest{} simulation is used as a reference climate state. 

The simulation \DPtest{} is explored and compared to \COtest{}. 
Both have exactly the same set of parameters, but \DPtest{} has an open \DP{}, while \COtest{} has a closed barrier in the southern ocean. 

\begin{figure}[tbh]
 \centering

\includegraphics[width=0.65\textwidth]{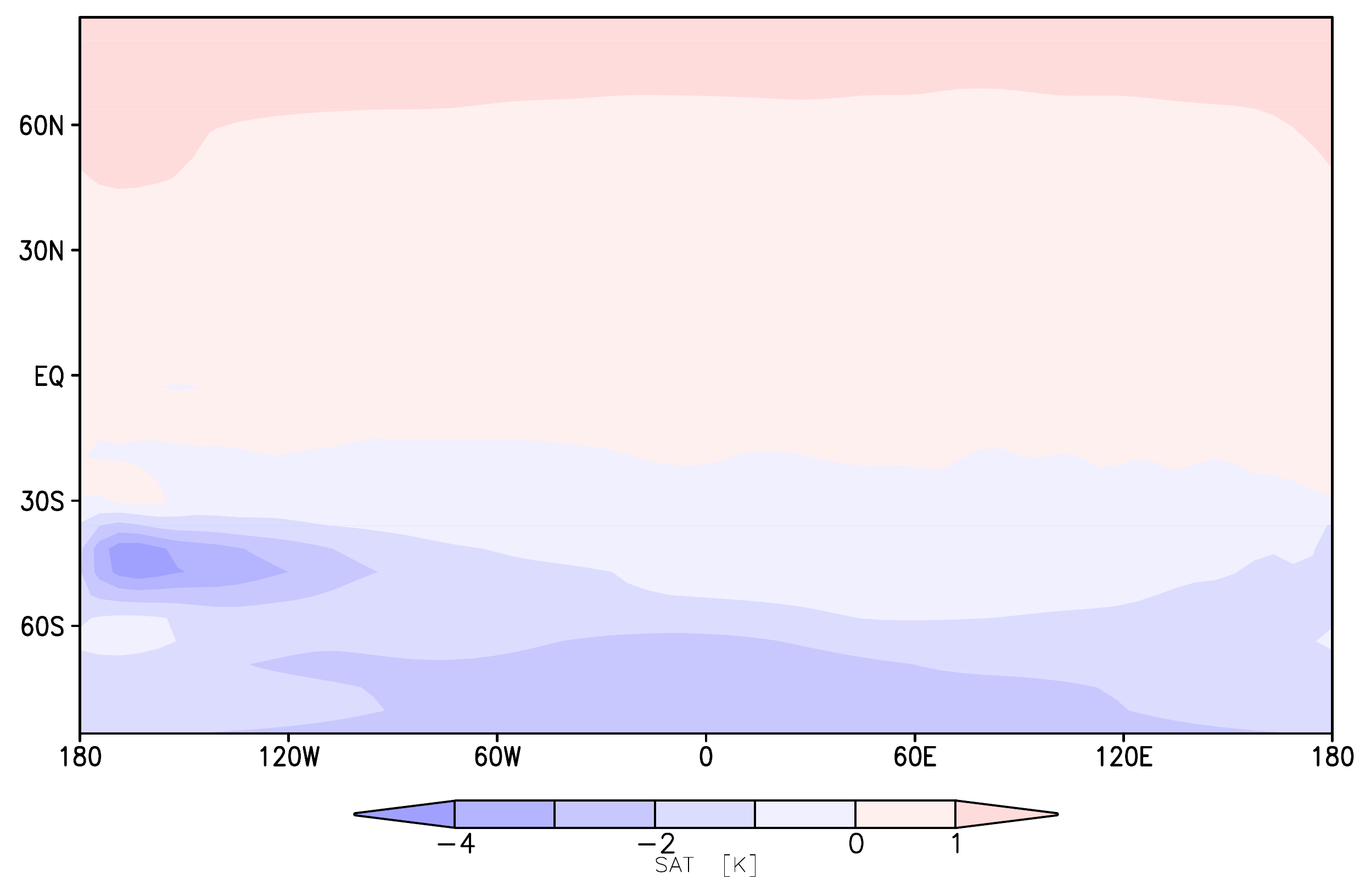}

\caption{Surface air temperature difference between the \DPtest{} and the \COtest{} climates}
\label{fig:DPtest-CO2test_SATdiff}
\end{figure}

The differences in the surface air temperature between \DPtest{} and \COtest{} are displayed in figure~\ref{fig:DPtest-CO2test_SATdiff}. 
The SAT difference shows a distinct anti-symmetric pattern about the equator. 
The northern hemisphere experiences a warming when the \DP{} opens up. 
The northern polar latitudes are warmed by over \unit{1}\kelvin{}. 
However, poleward of \unit{30}\degree{}S (where the \DP{} begins in this idealized set-up), the air close to the surface is cooled by \unit{1--4}\kelvin{}. 
The areas with the strongest cooling are located over the southern ocean east of the \DP{} opening and at the center of the continent. 

The cooling of the southern continent caused by the opening of the \DP{} is also not strong enough to explain the large temperature drop during the transition from \contr{} to \final{} alone. 
Even when added to the cooling accounted for by  the \CO{} decrease, there is hardly more than a temperature reduction of \unit{5--6}\kelvin{}. 
However, the differences between \contr{} and \final{} exceed \unit{20}\kelvin{} at the very high southern latitudes. 

Table~\ref{tab:snow} (in section~\ref{sec:S}) shows that even though the surface temperature decreases on the continent when the \DP{} opens up, there is hardly more snow and the albedo remains at approximately 0.4.

\subsection{The role of orbital parameters}\label{sec:S}

\begin{figure}[th!]
 \centering

\subfloat[\Stest{}--\COtest{}]{\includegraphics[width=0.6\textwidth]{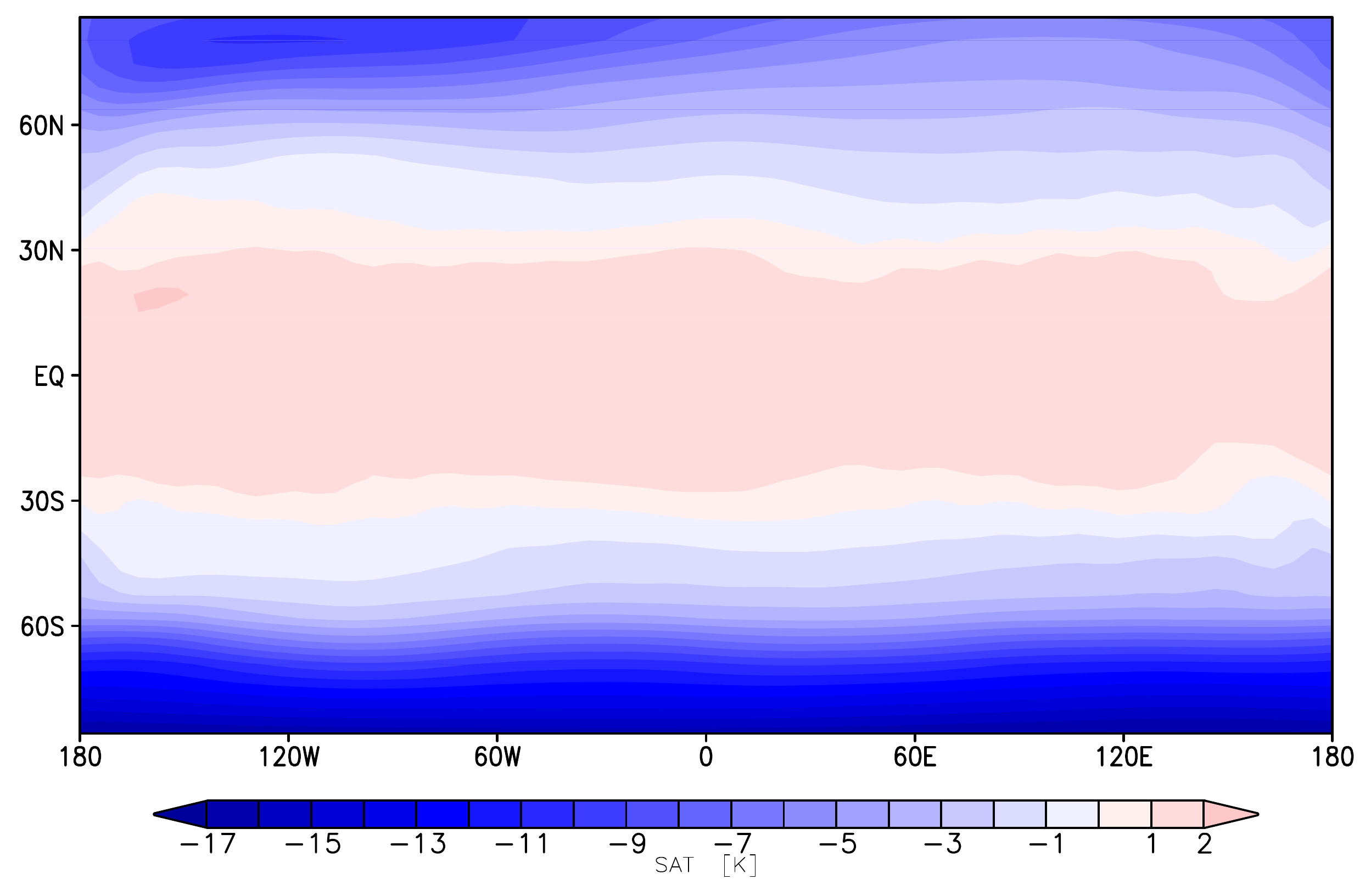}\label{fig:Stest-CO2test_SATdiff}}

\subfloat[\Stestelf{}--\COtest{}]{\includegraphics[width=0.6\textwidth]{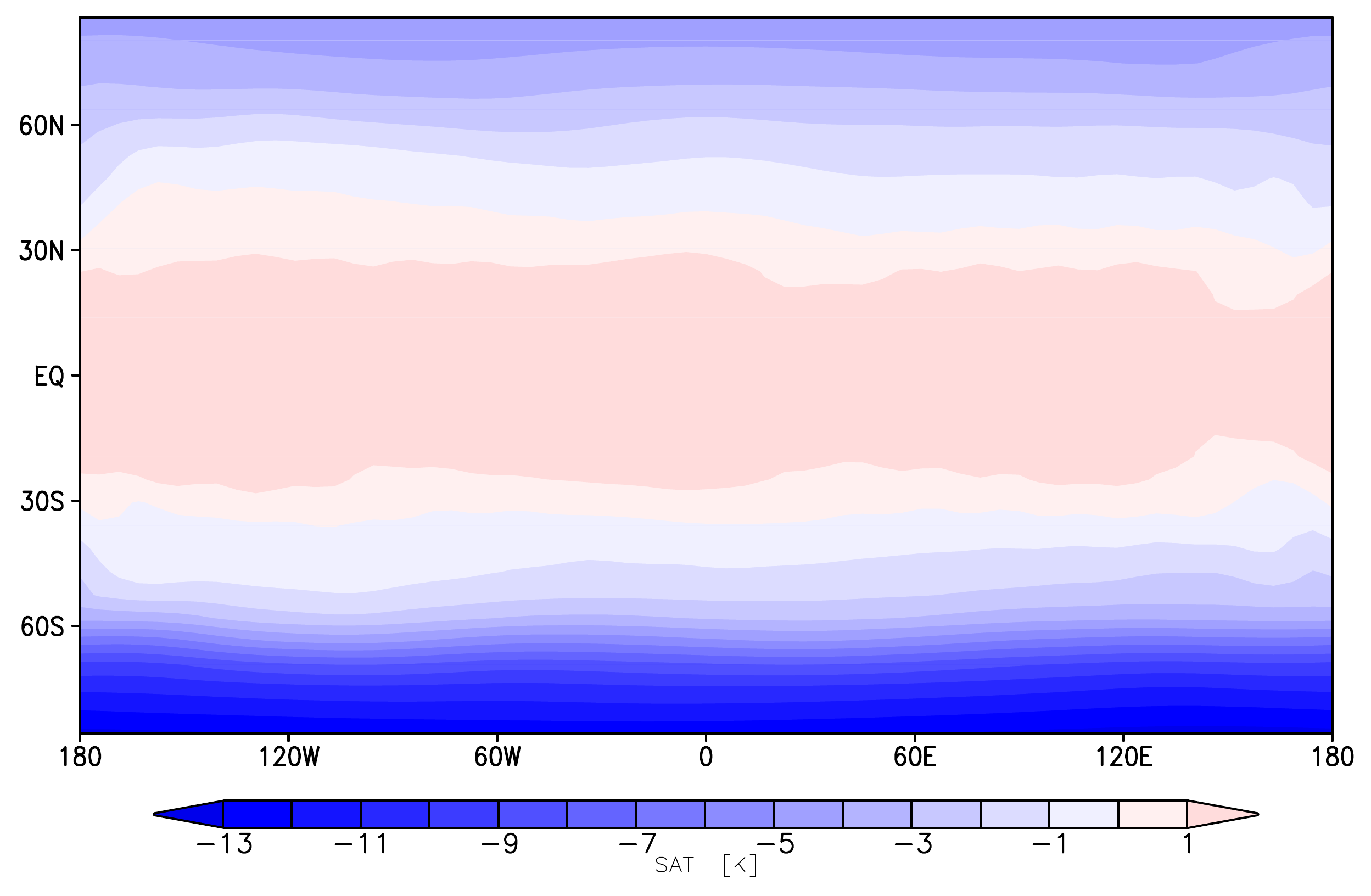}\label{fig:Stest11-CO2test_SATdiff}}

\subfloat[\Stestzz{}--\COtest{}]{\includegraphics[width=0.6\textwidth]{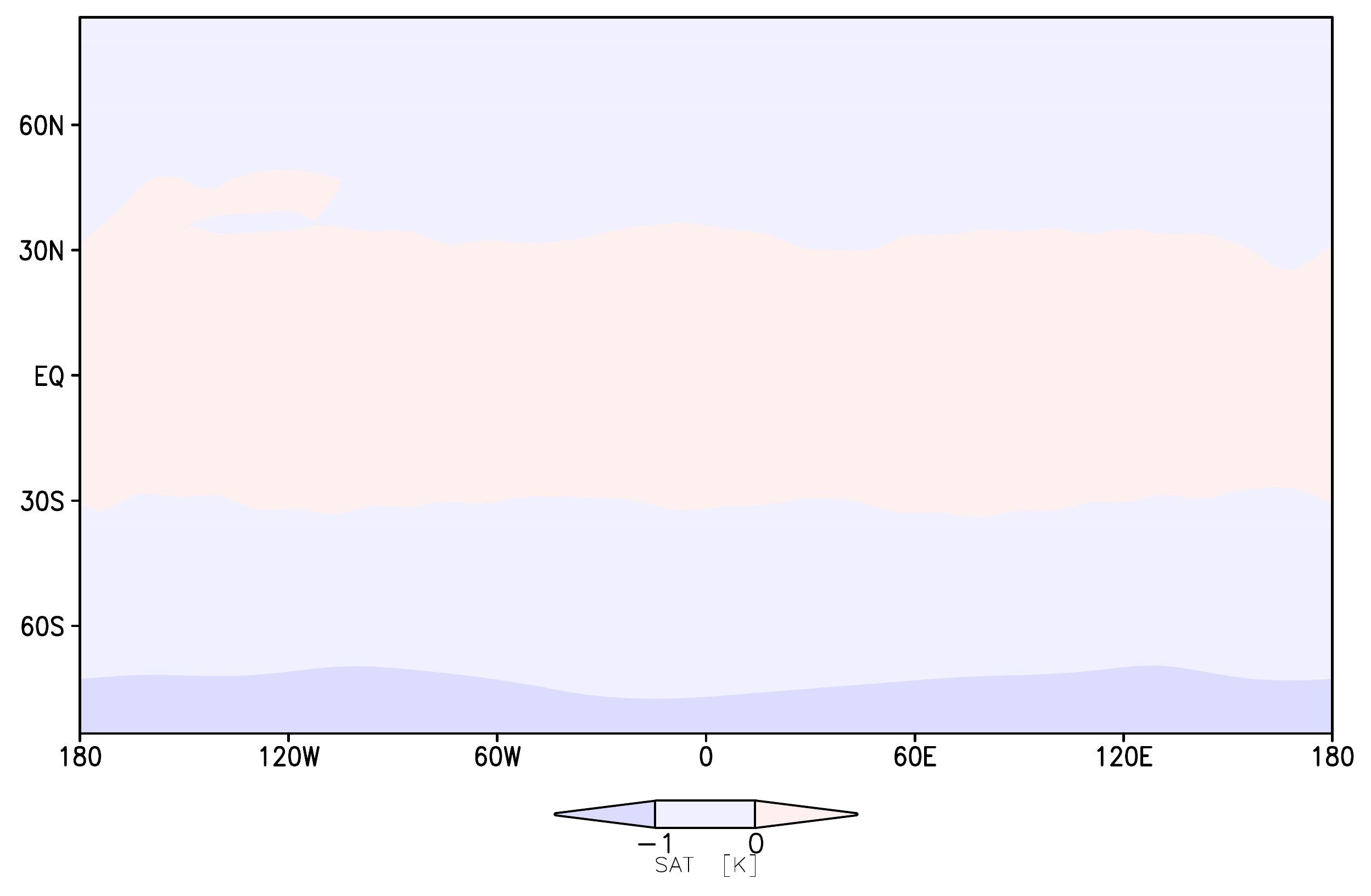}\label{fig:Stest22-CO2test_SATdiff}}

\caption{Surface air temperature difference between the \Stest{} and the \COtest{} climates}
\label{fig:Stestall-CO2test_SATdiff}
\end{figure}

\Stest{} has exactly the same set of parameters (and topographical set-up) as \COtest{}, except that the orbital parameters are changed to exclude a seasonal cycle. 
The SAT difference between \Stest{} and \COtest{} is shown in figure~\ref{fig:Stest-CO2test_SATdiff}. 
The seasonality causes temperature deviations which are almost symmteric about the equator: in the tropics and subtropics there is a warming (up to \unit{2}\kelvin{}) and the extra-tropical air is cooled. 
Most prominently, it can be observed that temperatures at the poles, and especially at the south pole, are reduced dramatically. 
At the north pole, the high latitudes experience a strong cooling east of the ocean basin boundary (more than \unit{10}\kelvin{}) and a moderate cooling (\unit{5--6}\kelvin{}) west of the boundary. 
The air above the southern continent is cooled almost uniformly and the temperature reduction exceeds \unit{15}\kelvin{} poleward of the continental boundary. 

When analyzing all three parameters (atmospheric \CO{}, opening of the \DP{}, and seasonality) alone, it can be concluded that the strongest cooling at the southern continent (during the transition from the \contr{} to the \final{} climate) takes place when the orbital parameters are changed to zero obliquity. 
The seasons are responsible for the melting periods in the austral summer (see figure~\ref{fig:final-contr_SATsnow}), so that the snow cover is substantially reduced, when orbital parameters allow a strong seasonal cycle. 

\afterpage{\clearpage}

Annual mean snow depth, snowfall, and snow melt averaged over the continental area are listed in table~\ref{tab:snow} for all sensitivity experiments. 
Furthermore, the surface albedo and the surface temperature are given. 

\begin{table}[b!]
\centering
\small
\begin{tabular}{llccccc}
 \toprule
&&\textbf{snow} & \textbf{snow-} & \textbf{snow} & \textbf{surface} & \textbf{surface} \\
&&\textbf{depth} & \textbf{fall} & \textbf{melt} & \textbf{albedo} & \textbf{temperature} \\
&&[\meter{}] & [\centi\meter\per{}month] & [\centi\meter\per{}month] & & [\kelvin{}] \\
\midrule
\textsc{present-} &  \textbf{\contr{}} & 0.06& 2.59& 2.09& 0.39& 262.9\\
\textsc{day} & \textbf{\COtest{}} & 0.07& 2.65& 2.08& 0.41& 259.0 \\
\textsc{obliquity} & \textbf{\DPtest{}} & 0.07 & 2.56 & 1.97 & 0.41 & 256.6 \\
\midrule
\textsc{reduced}&\textbf{\final{}} & 4.98& 2.66& 0.77& 0.58& 238.0\\
\textsc{obliquity}&\textbf{\Stest{}}& 3.95& 2.98& 1.11& 0.57& 242.0\\
&\textbf{\Stestelf{}}& 3.32 & 3.23 & 1.50 & 0.54 & 245.4 \\
&\textbf{\Stestzz{}}& 0.07 & 2.76 & 2.16 & 0.41 & 257.9 \\
 \bottomrule
\end{tabular}
 \caption{Annual mean values of snow depth, fall, and melting rate, as well as surface albedo and temperature for the different set-ups averaged over the continental area}
 \label{tab:snow}
\end{table}

Under perpetual equinoctial conditions (in \final{} and \Stest{}), the seasonal cycle is suppressed and the snowfall is higher than the melting rate all year long (see table~\ref{tab:snow}). 
The constant snow cover on the southern continent creates a snow--albedo feedback, which further lowers the temperatures in this area. 
For the cases with present-day orbital parameters (\contr{}, \COtest{}, and \DPtest{}), the continent is completey snow-free in austral summer, while snow can accumulate during austral winters (note that annual mean snowfall and snow melt do not completely compensate since sublimation has to be considered). 
The annual mean surface temperatures are higher for these three cases and the albedo over the continent decreases from approximately 0.6 to 0.4 with a seasonal cycle. 

\begin{figure}[bt]
 \centering

\includegraphics[width=0.65\textwidth]{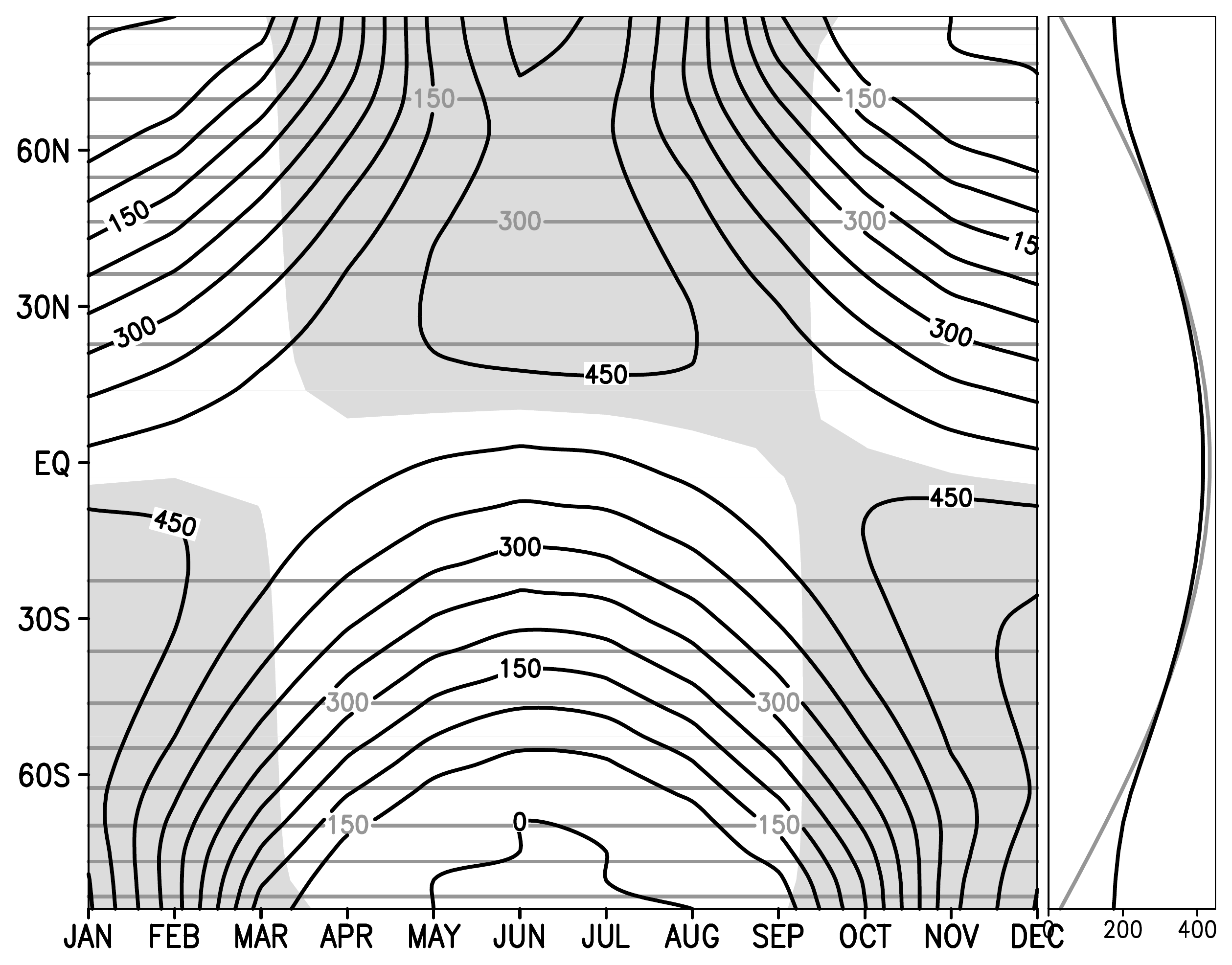}

\caption{Seasonal cycle (\textit{left}) and annual mean (\textit{right}) of the zonal mean insolation forcing for the cases with present-day seasonality (black contours: \contr{}, \COtest{}, and \DPtest{}) and the cases with perpetual equinoctial conditions (gray contours: \Stest{} and \final{}; gray shading indicates that the difference in insolation (\final{} minus \contr{}) is negative)}
\label{fig:insolation}
\end{figure}

The change in the orbital parameters (i.e.\ obliquities of approximately \unit{23.5}\degree{} and \unit{0}\degree{}, for details see section~\ref{sec:exp}) mainly causes differences in the insolation forcing. 
The seasonal cycle of the zonal mean incoming solar radiation (at the top of the atmosphere) is displayed in figure~\ref{fig:insolation}. 
The annual mean insolation is also shown (on the right). 
Figure~\ref{fig:insolation} reveals that especially the high latitudes are affected by the change in orbital parameters. 
Under the present-day seasonal cycle, the continental area at the south pole receives between \unit{0}\watt\per\squaren\meter{} and \unit{100}\watt\per\squaren\meter{} of incoming solar radiation during austral winter (May to August), while the insolation reaches values between \unit{450}\watt\per\squaren\meter{} and \unit{500}\watt\per\squaren\meter{} during the summer months (November to February). 
In contrast, under perpetual equinoctial conditions the solar forcing at the continent is always between \unit{50}\watt\per\squaren\meter{} and \unit{200}\watt\per\squaren\meter{}. 
At the southern continent the insolation during austral summer is higher in \contr{} than in \final{}, however, during austral winter there is less insolation at high latitudes. 
While the tropics receive slightly more insolation in the annual mean under perpetual equinoctial conditions, the poles show a great insolation deficit compared to the present-day conditions. 
This forcing is even more important than the snow--albedo feedback (over the continent the snow--albedo feedback causes differences between \contr{} and \final{} of approximately \unit{20}\watt\per\squaren\meter{} to \unit{50}\watt\per\squaren\meter{}, while the changed insolation causes differences between \unit{30}\watt\per\squaren\meter{} and \unit{140}\watt\per\squaren\meter{}). 

Setting the obliquity to zero and comparing it to a simulation with an obliquity of \unit{23.5}\degree{} is an extreme case, even though the aim of this study is not to realistically simulate the onset of the Antarctic glaciation, but to analyze the general mechanisms. 
To see how much the results obtained here depend on the chosen obliquity settings, we study two further simulations with obliquities between zero and present-day. 
\Stestelf{} and \Stestzz{} have identical set-ups as \Stest{}, except that the obliquity is set to \unit{11.5}\degree{}, i.e.\ \unit{22.5}\degree{}. 

In figure~\ref{fig:Stestall-CO2test_SATdiff}, the SAT differences between the three \Stest{} set-ups and \COtest{} are shown. 
The overall pattern is the same in all three set-ups: there is a warming in the tropics and a cooling at high latitudes. 
However, the magnitude of the warming and cooling patterns differ. 
The strongest temperature differences (up to \unit{-17}\kelvin{}) can be observed in \Stest{} (with an obliquity of \unit{0}\degree{}). 
\Stestelf{} shows a more moderate cooling at northern high latitudes, but the southern continent cools still very strongly (up to \unit{-13}\kelvin{}). 
Even though there are obliquity changes by \unit{11.5}\degree{}, \Stest{} and \Stestelf{} give relatively similar results: the southern continent is very cold (\unit{242.0}\kelvin{} i.e.\ \unit{245.4}\kelvin{} on average, see table~\ref{tab:snow}), there is constant snow accumulation and a high surface albedo. 

When the obliquity is reduced by only \unit{1}\degree{},  in \Stestzz{}, only a slight cooling of \unit{1--2}\kelvin{} can be observed at the southern polar continent. 
Furthermore, snow parameters, surface albedo and surface temperature are rather comparable to \DPtest{} or \COtest{} (see table~\ref{tab:snow}). 

It can be concluded that when altering the obliquity only slightly, all three forcing mechanisms, opening of the \DP{}, decreasing \CO{}, and reducing obliquity, have effects of similar importance for the climate of the southern polar continent. 
A cooling still takes place by about \unit{7.5}\kelvin{} (adding all three effects), but there is still a pronounced melting period in austral summer, so that snow does not accumulate on the continent. 
However, reducing the obliquity to \unit{11.5}\degree{} is already sufficient to create very cold conditions (a cooling of about \unit{21}\kelvin{} with all three forcings included) on the southern polar continent, where snow is able to accumulate constantly over the year. 

\subsection{Combined effects}

To analyze to what degree the three forcing factors can be added linearly to obtain the \final{} climate state, the factor separation after \citet{Stein1993} has been applied to the data. 
The factor separation is a method for computing interactions among various factors influencing, in our case, the surface air temperature averaged over the continental area. 

\begin{figure}[tbh]
 \centering

\includegraphics[width=0.5\textwidth]{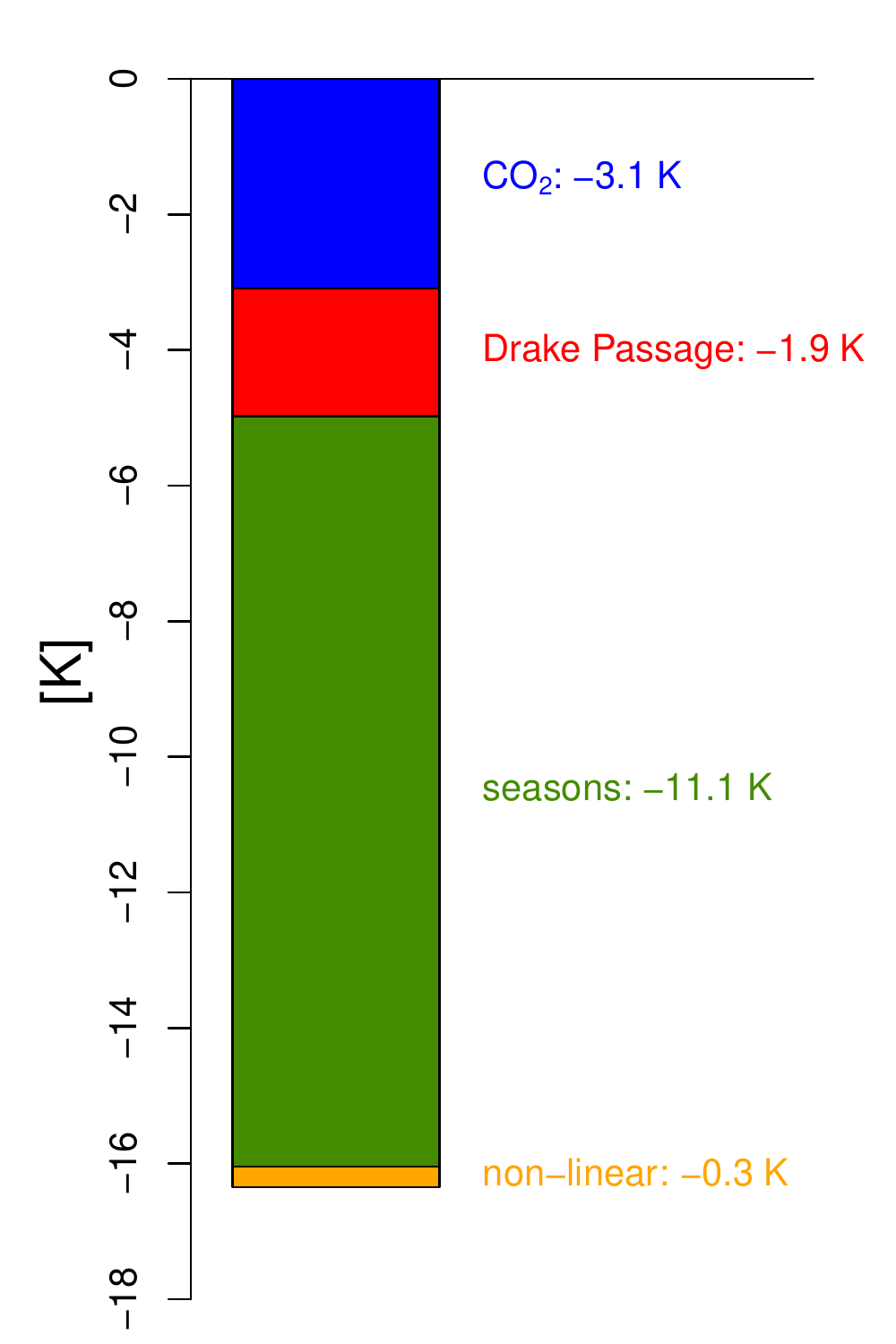}

\caption{Factor separation of the forcings controlling the cooling of the southern continent: temperature differences caused by \CO{} alone, the \DP{} alone, the seasonal cycle alone, all non-linear effects together.}
\label{fig:SAT_factor}
\end{figure}

Figure~\ref{fig:SAT_factor} shows the roles of \CO{}, the \DP{}, and the orbital parameters (setting the obliquity to zero) alone, as well as all combined (non-linear) effects together. 
It can be observed that the individual forcings are much more important for the temperature change on the southern continent than nonlinear effects, which are responsible for only approximately \unit{2}\%{} of the cooling. 

\section{Discussion and conclusion}
\label{sec:conclusion}

Coupled idealized simulations are presented to explore the roles of atmospheric \CO{} decrease, opening of ocean gateways, and changing orbital parameters for a fictional climate transition, that resembles the onset of the Antarctic glaciation during the Eocene--Oligocene transition. 
An aquaplanet, which includes a circular continent at the south pole and a meridional barrier in the ocean, is applied for the sensitivity experiments. 
While idealized set-ups like aquaplanets are not a tool for producing realistic climate simulations in a quantitative sense, they can significantly contribute to our understanding of the climate system in its most elemental form \citep[see for example][]{Smith2006,Marshall2007,Enderton2009,Ferreira2010,Ferreira2011,Dahms2011,Hertwig2015}. 

Comparing the respective simulations (\contr{} and \final{}), we show that the combined effects of decreasing \CO{}, opening of the \DP{}, and switching off the annual cycle can indeed trigger a climate transition, in particular at the southern polar continent. 
This is documented, for example, by changes in temperature, snowfall (snow accumulation), and meridional heat transport. 

At the south pole the temperature decrease is by far the strongest, but the climate also cools in the global mean. 
The surface air temperature over the continent decreases by more than \unit{20}\kelvin{} between \contr{} and \final{}. 
The simulated temperature reduction exceeds values from sedimentary records from the Eocene--Oligocene transition \citep[see for example][]{Zachos2008}. 
However, it is not expected that exact temperature changes are reproduced, since the climate transition simulated here only resembles the transition during the glaciation of Antarctica due to the degree of idealization. 

The total poleward heat transport in the \final{} climate exceeds the transport in the \contr{} simulation. 
Since the obliquity and the mean hemispheric albedo are altered during the transition, the concept of a constant total meridional energy transport \citep{Stone1978} does not hold anymore. 
The ocean heat transport (OHT) is very large in the northern hemisphere of the \final{} climate. 
While peak values are located in the subtropics, a very strong OHT can still be observed at high latitudes. 
Furthermore, in the southern hemisphere of the \final{} climate and in both hemispheres of the \contr{} climate, the OHT is relatively constant with values up to \unit{1}\peta\watt{}. 
The OHT differs from present-day estimates \citep[see for example][]{Trenberth2001a} as well as from the one seen in the \DP{} simulation by \citet{Enderton2009}, which also peaks in the subtropics, but is already very weak poleward of the mid-latitudes. 

Between the \contr{} and \final{} set-up, all three forcing mechanisms differ: atmospheric \CO{} is reduced, the \DP{} is opened, and the seasonal cycle is switched off. 
To determine which of these forcings has the greatest impact on the global climate and on the climate transition of the southern polar continent, several sensitivity experiments are analyzed separately:

\begin{itemize}
 \item \textbf{Declining atmospheric CO$_\mathbf{2}$:} The \CO{} decrease has a great cooling effect on the global scale. 
  The climate with higher \CO{} values (\contr{}) is characterized by warm temperatures, shallow meridional temperature gradients, and a weak circulation in the atmosphere. 
  When comparing this to a climate state with the same parameters but lower \CO{} values (\COtest{}), the temperatures are globally reduced, but the reduction over the continent (\unit{2--3}\kelvin{}) is not strong enough to explain the huge temperature reduction from \contr{} to \final{} (up to \unit{20}\kelvin{}). 
  Most notably, there is still a pronounced melting period which causes completely snow-free summer months on the southern polar continent. 
 \item \textbf{Opening of the \DP{}:} The oceanic circulation is most strongly affected by the opening of the \DP{}. 
  When the southern ocean gateway opens up, a very strong circumpolar current develops. 
  As a consequence, the southward OHT is almost completely inhibited in the area of the \DP{} and the southern ocean becomes thermally isolated and very cold. 
  Thus, the temperatures above the southern continent indeed decline when the \DP{} is opened (while the latitudes northward of the passage are warmed). 
  However, when the \DP{} is opened and all other forcings remain the same (\DPtest{}), the temperatures over the continent reduce by approximately \unit{2--3}\kelvin{}, which explains only about \unit{10}\%{} of the overall cooling from \contr{} to \final{}. 
  Furthermore, the melting period causing the snow-free summer months is still present.  
 \item \textbf{Changing orbital parameters:} Under perpetual equinoctial conditions the low latitudes receive more insolation and experience an overall warming, while the poles become much colder when seasons are excluded. 
  Under present-day seasonality, the poles receive ample insolation during polar summers and hardly/no insolation during polar winters. 
  Thus, the high latitudes are more strongly affected by the change in orbital parameters than low and mid-latitudes. 
  When the seasonal cycle is suppressed (and all other parameters remain the same, \Stest{}), the temperatures over the southern polar continent decline by up to \unit{15}\kelvin{}. 
  Hence, changing orbital parameters explain the greatest part of the massive temperature reduction from the \contr{} to the \final{} climate state. 
  Because of the missing seasonality there is no summertime melting and snow accumulates all year long. 
  Reducing the obliquity to \unit{11.5}\degree{} (\Stestelf{}) instead of \unit{0}\degree{} is already sufficient to cause very strong cooling at the southern continent and to suppress the summertime melting of snow. 
  However, changing the obliquity by only \unit{1}\degree{} (\Stestzz{}) is not enough to suppress the summer melting periods and the continent is cooled by \unit{1--2}\kelvin{} only, which is comparable to \COtest{} and \DPtest{}. 
\end{itemize}

In summary, all three forcings cool the climate of the southern polar continent. 
However, even though declining atmospheric \CO{} and the opening of the \DP{} reduce the temperatures over the southern continent, the cooling caused by the changing orbital parameters is considerably stronger when setting the obliquity to zero or to \unit{11.5}\degree{}. 
For an obliquity of \unit{22.5}\degree{}, all three forcings are of the same order of magnitude for the cooling of the southern polar continent. 
Furthermore, the summertime melting period which prevents the constant accumulation of snow is only suppressed when orbital parameters are altered strongly (in \Stest{} or \Stestelf{}). 

Studying the transition from \contr{} to \final{}, the oppressed seasonality causes the greatest differences. 
When the seasonal cycle is suppressed, temperatures on the continent cool dramatically and snow accumulates persistently. 
However, even though changes in \CO{}, ocean gateways, and orbital parameters are difficult to compare, the alteration in the obliquity might be too extreme compared to the other forcing factors (with a change by approximately \unit{23.5}\degree{}). 
While setting the obliquity to \unit{11.5}\degree{} is already sufficient to create a climate transition like the one between \contr{} and \final{}, the climate transition with an obliquity of \unit{22.5}\degree{} is not very pronounced (colder temperatures, but no snow accumulation). 

There are numerous studies dealing with the onset of the Antarctic glaciation aiming to determine its main forcing mechanisms. 
The focus is placed on the three main hypotheses: declining atmospheric \CO{}, the opening of the Drake Passage, and varying orbital parameters. 

Most estimates of atmospheric \CO{} mixing ratios in the early Cenozoic range between $2\times$ and $5\times$ of the pre-industrial values and during the transition atmospheric \CO{} is believed to have dropped to nearly present-day values \citep{Pagani1999,Pearson2000}. 
\citet{DeConto2003a,DeConto2003} analyze simulations with decreasing atmospheric \CO{} and find a threshold for the formation of an ice-cap between $3\times$ and $2\times$\CO{} (pre-industrial). 
They identify declining atmospheric \CO{} as the main forcing for the Cenozoic climate change. 
In contrast, in the climate transition modeled in this study, decreasing atmospheric \CO{} is only part of the forcing that cools the southern polar continent. 
Furthermore, the effect of declining greenhouse gas concentrations rather results in a global cooling than in a local climate change. 
In the case of changing the obliquity only by \unit{1}\degree{} (to \unit{22.5}\degree{}), which is in the order of magnitude of the experiments by \citet{DeConto2003a,DeConto2003}, all three forcings have a similar impact on the climate of the southern polar continent. 

The differences between our results and the findings of \citet{DeConto2003a,DeConto2003} can also be explained with the different experimental set-up. 
\citet{DeConto2003a} observe that when the atmospheric \CO{} values decrease to the threshold between $3\times$ and $2\times$ pre-industrial \CO{}, ice caps can grow fast because height--mass balance feedbacks are initiated. 
Since a land ice model is not included in our set-up, these feedbacks are excluded. 
Furthermore, only one continent exists in our idealized planet. 
In the model of \citet{DeConto2003a,DeConto2003} snow can also accumulate on other continents when decreasing \CO{} cools the global climate affecting, for example, the planetary albedo and adding a non-linear response, which is not included in this idealized set-up. 

\citet{DeConto2003a,DeConto2003} also analyze the role of orbital parameters and the opening of the Drake Passage for the onset of the Antarctic glaciation. 
While they conclude that the role of the Drake Passage is rather minor, they find that the orbital parameters do indeed affect the onset of the glaciation. 
Their threshold around $3\times$ pre-industrial \CO{} can only establish a continental ice sheet if the orbital parameters favor cool austral summers. 
However, once the atmospheric \CO{} declines further, the AIS becomes almost insensitive to the orbital forcing. 

A direct comparison between our results and \citet{DeConto2003a,DeConto2003} is not possible, since they apply realistic topographic boundary conditions and reduce atmospheric \CO{} in multiple (smaller) steps. 
Furthermore, they only slightly reduce the seasonal cycle, comparable to our \Stestzz{} set-up. 
However, it should be noted that in our set-up we do not attempt to realistically model the build-up of the Antarctic Ice Sheet, but are interested in general mechanisms. 
An idealized land--sea configuration can be very useful for reducing the climate system to its most elemental features and, thus, help to identify the main characteristics of the governing processes. 
However, while the idealized model contains the same basic processes as more complex systems, one has to be careful when interpreting the results and relating them to reality. 

\citet[][for example]{Cristini2012} present a senstivity study aiming to understand if and how the opening of the Drake Passage served as a forcing factor for the Antarctic climate transition. 
In their simulations the SAT averaged over the Antarctic continent cools only by \unit{0.4}\kelvin{} when the Drake Passage is opened, but they observe an additional growth of the Antarctic Ice Sheet volume of about \unit{20}\%{} and conclude that the opening of the Drake Passage contributed substantially to the Antarctic glaciation. 
However, \citet{Cristini2012} do not modify atmospheric \CO{} concentrations or orbital parameters and, therefore, could not answer the question about the importance of these possible mechanisms. 
In this study, the opening of the idealized ocean gateway cools the climate of the southern continent by about \unit{2--3}\kelvin{}, which is the same order of magnitude as found in the idealized \DP{} study by \citet{Toggweiler2000}. 

However, one has to be careful when comparing the meridional ocean barrier of the idealized set-ups to the Drake Passage of the era during the Antarctic glaciation, since \citet{Zhang2010a} found that the climate impacts of the Drake Passage strongly depend on the land--sea configuration and the passage geometry. 
Furthermore, the aquaplanet ocean has a flat sea floor and no additional continents. 
Even though in the presence of the meridional barrier subpolar and subtropical gyres develop that resemble the ocean circulation of the Atlantic or Pacific, the ocean circulation deviates greatly from the one found in models with a rather realistic topographical set-up. 

Still, idealized model set-ups are useful for isolating and identifying the individual effects. 
Furthermore, the zonal mean atmospheric state can to a large degree be approximated by an aquaplanet \citep[see for example][who analyzed the mean state and the variability of a pure aquaplanet]{Marshall2007,Ferreira2010,Ferreira2011,Hertwig2015}. 
An idealized set-up similar to this study but without the polar continent is investigated by \citet{Smith2006} and \citet{Enderton2009}. 
Their respective experiments with an idealized \DP{} most resembles our present-day climate (compared to the other set-ups with or without a meridional ocean barrier). 

In the idealized set-up of this study, a change in obliquity to \unit{0}\degree{} or \unit{11.5}\degree{} alone is sufficient to trigger a massive cooling of the southern polar continent, even though the opening of the \DP{} and declining \CO{} also contribute to the cooling. 
While varying setting the obliquity to \unit{0}\degree{} is by far the most important forcing for the climate transition from \contr{} to \final{} simulated in this study, one cannot conclude that they are also the main forcing for the onset of the Antarctic glaciation. 
We show that the climate transition caused by the orbital parameter change is dependent on the obliquity. 
Even though, considering the results of this study, the role of the varying seasonality for past climate changes may be re-evaluated and more emphasis could be placed on the orbital parameters in future studies dealing with the Antarctic glaciation. 

While there is no land ice model, a thermodynamic sea ice model is coupled to the Planet Simulator--LSG model. 
However, sea ice does not form in the experiments described here, which is due to the diffferent climate conditions on the aquaplent. 
This has to be considered when interpreting the results and trying to link them to reality, since the absence of sea ice may alter temperatures and precipitation over the southern ocean. 
Furthermore, sea ice is important for the build up of land ice, which is not included in this set-up. 

As a next step, the degree of idealization of this set-up could be reduced to make the results better comparable to other model studies and data records concerning the Antarctic glaciation. 
The inclusion of an ice sheet model could have additional feedbacks, which are not included in these simulations. 
Today, the height of the Antarctic Ice Sheet exceeds \unit{3,000}\meter{} over large areas \citep{Lythe2001}. 
At this high elevation, there is a cooling effect caused by the uplift of air. 
This positive feedback (build-up of an ice sheet, higher elevation, uplift, additional cooling) is missing in our simulations since the idealized continent does not contain topographical features nor a model for land ice. 

Further simulations are called for to challenge or strengthen the results obtained in this study. 
These include simulations with a fully coupled model including a land ice model. 
Also further sensitivity studies could complement this set of simulations in which more than one parameter at the time could be altered. 
A more thorough analysis of the role of the topographical set-up could be conducted by a stepwise transition to a more realistic continental distribution, which could include a second passage in the tropics (representing the Panama Passage), a second oceanic barrier with a passage in the southern ocean (the Tasman Passage), topographical features on the ocean floor, and more (idealized) continents. 
The role of declining \CO{} could be analyzed in more detail by following the line of \citet{DeConto2003a,DeConto2003} and reducing the atmospheric \CO{} level in multiple (smaller) steps. 
Transient \CO{} changes could also affect the results \citep[see for example][]{Bordi2012}. 

Most importantly, the role of the varying orbital parameters has to be analyzed in more detail, since they appear to be essential for the climate transition simulated in this study. 
Additional sensitivity studies with smaller changes in the obliquity would be a natural next step to find at which obliquity the orbital parameters become the dominant forcing. 

A final comment appears in order. 
This idealized experimental setting in the Planet Simulator environment is embedded in a wide range of studies: vegetation impact associated with regional tipping point analyses and a comparison of a green versus desert globe \citep{Fraedrich2005c,Bathiany2012}, a world with and without Greenland \citep{Junge2005}, global scale hysteresis experiments with statically changing radiation and transient \CO{}-change \citep{Fraedrich2012}, idealized aqua-planet circulations related to double jet dynamics modulated by the stratosphere and the tropical double ITCZ circulation \citep{Bordi2007,Dahms2011}. 
Here we have presented a first extension of the spectrum of idealized climate dynamic experiments by including a simple ocean model.

\section*{Acknowledgments}
EH and KF acknowledge and are grateful for the support by a Max Planck Fellowship.

\bibliographystyle{apalike}      

\end{document}